\documentclass[review]{elsarticle}
\usepackage{interval}
\usepackage{lineno,hyperref}
\usepackage{graphicx}
\usepackage[bottom]{footmisc}
\usepackage{amsmath,amssymb}
\usepackage{amssymb}
\usepackage{algorithm}
\usepackage{algorithmic}
\usepackage{appendix}
\usepackage{mathtools}
\usepackage{stackengine}
\usepackage[section]{placeins}
\usepackage{longtable}
\usepackage{enumitem}
\usepackage{multirow}
\usepackage{subfig}
\usepackage{hhline}
\usepackage{breqn}
\usepackage{xcolor}
\newcommand{\minimize}[2]{\ensuremath{\underset{\substack{{#1}}}{\textrm{minimize}}\;\;#2 }}

\usepackage{threeparttablex}

\usepackage{color}
\newcommand{\cred}{\textcolor{red}}

\modulolinenumbers[5]

\biboptions{authoryear}

\begin{document}
\begin{frontmatter}
\begin{titlepage}
   \begin{center}
       \vspace*{1cm}
       \LARGE
\textbf{SuperDeConFuse: A Supervised Deep Convolutional Transform based Fusion Framework for Financial Trading Systems}

\vspace{1.5cm}
\large\textbf{Pooja Gupta}\\
\small Indraprastha Institute of Information Technology, Delhi, India \\
\small poojag@iiitd.ac.in\\
\vspace{0.5cm}
\large\textbf{Angshul Majumdar}\\
\small Indraprastha Institute of Information Technology, Delhi, India \\
\small angshul@iiitd.ac.in\\
\vspace{0.5cm}
\large\textbf{Emilie Chouzenoux}\\
\small CVN, Inria Saclay, CentraleSup\'elec, Gif-sur-Yvette, France \\
\small emilie.chouzenoux@centralesupelec.fr\\
\vspace{0.5cm}
\large\textbf{Giovanni Chierchia}\\
\small LIGM, Universit\'e Gustave Eiffel, CNRS, ESIEE Paris, Noisy-le-Grand, France \\
\small giovanni.chierchia@esiee.fr
% \author[mymainaddress]{Giovanni Chierchia\corref{mycorrespondingauthor}}
% \cortext[mycorrespondingauthor]{Corresponding author}
% \ead{giovanni.chierchia@esiee.fr}

% %% or include affiliations in footnotes:
% \author[mymainaddress,mysecondaryaddress]{Angshul Majumdar}
% \ead{angshul@iiitd.ac.in}

% \author[mymainaddress2]{Emilie Chouzenoux}
% \ead{emilie.chouzenoux@centralesupelec.fr}
% \author[mymainaddress3]{Giovanni Chierchia}
% \ead{giovanni.chierchia@esiee.fr}

% \address[mymainaddress]{Indraprastha Institute of Information Technology, Delhi, India}
% \address[mysecondaryaddress]{TCS Research, Kolkata, India}
% \address[mymainaddress2]{CVN, Inria Saclay, CentraleSup\'elec, Gif-sur-Yvette, France}
% \address[mymainaddress3]{LIGM, Universit\'e Gustave Eiffel, CNRS, ESIEE Paris, Noisy-le-Grand, France}
\end{center}
\end{titlepage}

\begin{abstract}
This work proposes a supervised multi-channel time-series learning framework for financial stock trading.
Although many deep learning models have recently been proposed in this domain, most of them treat the stock trading time-series data as 2-D image data, whereas its true nature is 1-D time-series data. Since the stock trading systems are multi-channel data, many existing techniques treating them as 1-D time-series data are not suggestive of any technique to effectively fusion the information carried by the multiple channels. To contribute towards both of these shortcomings, we propose an end-to-end supervised learning framework inspired by the previously established (unsupervised) convolution transform learning framework. Our approach consists of processing the data channels through separate 1-D convolution layers, then fusing the outputs with a series of fully-connected layers, and finally applying a softmax classification layer. The peculiarity of our framework, that we call \emph{SuperDeConFuse} (SDCF), is that we remove the nonlinear activation located between the multi-channel convolution layers and the fully-connected layers, as well as the one located between the latter and the output layer. We compensate for this removal by introducing a suitable regularization on the aforementioned layer outputs and filters during the training phase. Specifically, we apply a logarithm determinant regularization on the layer filters to break symmetry and force diversity in the learnt transforms, whereas we enforce the non-negativity constraint on the layer outputs to mitigate the issue of dead neurons. This results in the effective learning of a richer set of features and filters with respect to a standard convolutional neural network. Numerical experiments confirm that the proposed model yields considerably better results than state-of-the-art deep learning techniques for the real-world problem of stock trading.
\end{abstract}

\begin{keyword}
information fusion, deep learning, convolution, transform learning, stock trading.
\end{keyword}

\end{frontmatter}

% \linenumbers

\section{Introduction}
Financial time series forecasting, and particularly stock price forecasting, requires to determine the future value of a company’s stock or any other form of a financial instrument traded on exchange as per the company. It plays a significant role in trading strategies to identify opportunities to buy and sell a stock and this process is known as stock trading. This future movement prediction of stock could capitulate the significant profit. 

However, the problem of stock trading has been one of the most difficult problems for the researchers in finance data processing, and speculators. Struggles are mainly due to the uncertainties and noises of the samples. These samples are generated as a consequence of historical market behaviors. But their generation is also affected by other factors such as macroeconomy and investor feelings, hence it is not only dependent on historical information \citep{ref1}. Two famous hypotheses emphasize how difficult it is to accurately predict a stock price. First, the efficient market hypothesis introduced in \citep{ref12} states that the current price of an asset always reflects all previous information available for it instantly. Second, the random-walk hypothesis \citep{ref13} claims that stock price changes independently from its history. In other words, tomorrow’s price will only depend on tomorrow's information regardless of today’s price. Hence, automating the prediction of stock trends/movements is a very challenging task.

In past works, feature engineering played a key role in the prediction process. Features were extracted from the original stock data using technical analysis/indicators, which are in general used for analysing the stock market data. Traditional statistical methods such as linear regression, autoregressive moving average (ARMA), and GARCH, were much beneficial for financial time series forecasting due to their interpretability. These statistical models were thus used on the extracted features, processed using technical indicators related to historical data for future value prediction \citep{ref3}. Previous works have also used the extracted features as input to machine learning models like Naive Bayes (NB), Logistic Regression (LR), Random Forest (RF), and k-nearest neighbors (kNN) \citep{ref4,ref5,BALLINGS20157046}. 

In the last decade, deep learning based models/techniques have gained attention in multiple domains, and financial stock trading is one such domain. Owing to the success of Convolutional Neural Networks (CNNs), there are previous studies that have used this model for the future prediction of the stock value. In \citep{ref1}, the features extracted using technical indicators for stock data are fed as a 2-D ‘‘image’’ matrix to the CNN, where each column represents shifted windows of the data. The work \citep{ref2} utilizes long short-term memory (LSTM), deemed most suitable for time series analysis as they were supposed to mimic memory, and CNN for the stock trading task. However, it is likely that the most natural and thus efficient way to process time-series is to consider its original form as 1-D data rather than a 2-D matrix. It is worth mentioning that, up to our knowledge, despite its multi-channel form, the problem of financial stock trading has been rarely treated as a fusion problem. We can only mention \citep{ref6,ref7} where a fusion framework is proposed, but only at the feature level rather than at the raw level.  

In this work, motivated by the success of CNNs, we propose an end-to-end supervised fusion framework for multi-channel time-series based financial trading systems that makes use of our recently introduced convolutional transform learning (CTL) approach \citep{ref36}. We call this framework -\emph{SuperDeConFuse} (SDCF).\footnote{https://github.com/pooja290992/SuperDeConFuse.git} Our framework has the following contributions:
\begin{itemize}
    \item It is an end-to-end framework that treats the multi-channel time-series stock data as univariate data corresponding to every channel, thus overcoming both the aforementioned issues present in the previous works solving the problem of stock trading. 
    \item It promotes the learning of unique filters and hence a richer set of features, that was not guaranteed with CNNs, due to a ``logarithm determinant" penalty applied to the transforms/filters. 
    \item A non-negativity constraint on coefficients/features mitigates the dead neurons issue by removing the nonlinear activation of the fully-connected layers and the last convolution layer.
\end{itemize}

% In this work, we propose a supervised fusion framework that treats the multi-channel time-series stock data as univariate (1-D) data corresponding to every channel, thus overcoming both the aforementioned issues to solve the problem of stock trading. Motivated by the success of 1-D CNNs in financial data processing, we propose to make use of our recently introduced convolutional transform learning (CTL) approach \citep{ref36}.  We call this framework -\emph{SuperDeConFuse} (SDCF).\footnote{https://github.com/pooja290992/SuperDeConFuse.git} Our framework promotes the learning of unique features that is not easy to guarantee with CNNs. Additionally, a non-negativity constraint allows us to mitigate the dead neurons issue by removing the nonlinear activation of the fully-connected layers and the last convolution layer.  

The remainder of this paper is organized as follows. Section 2 summarizes related works in the field of machine learning and deep learning that have been proposed for solving the stock trading problem/ stock market prediction. 

Since our work focuses on a supervised multi-channel fusion framework, we will also review recent machine learning approaches for information fusion. Section 3 introduces the details of our proposed SuperDeConFuse (SDCF) approach, the mathematical tools involved and the training strategy that is retained. Section 4 discusses the considered dataset, data labeling, data preprocessing and the training methodology used. Section 5 provides the experimental results. Finally, Section 6 concludes this work.

\section{Literature Review}
\subsection{Financial stock data analysis}

In literature, different methodologies have been applied to the stock data for predicting future trading strategies (eg, buy and sell decisions). These include statistical methods, machine learning algorithms like Support Vector Machine (SVM) and Artificial Neural Networks (ANN), feature extraction approaches, deep learning models (eg, CNN, LSTM), that we briefly review in this section.

Statistical methods are probably the methods among others that are universally used for the prediction of financial stock trading strategies. In particular, many studies rely on the use of sequential statistical models, such as ARMA \citep{ref8}, ARCH \citep{ref9}, GARCH \citep{ref10} and \citep{ref11}, Kalman filter \citep{ref15}. 

Feature-based techniques are also considered as state-of-the-art. Technical indicators like Exponential moving average (EMA), Moving average convergence and divergence (MACD), Williams \%R, etc. have been used in past studies to extract the features from the data. Text mining can be used to process financial analysis from newspapers \citep{ref14}.  
%In  \citep{ref14}, a mathematical model based on sparse matrix factorization is applied over the features extracted using text mining on the Wall Street Journal articles for the prediction. %Kalman filter methods are also the mathematical methods that are generally used to analyze and forecast financial stock time series data in literature }. 
% 
The features are then used as input to machine learning models, for example, SVM, ANN, kNN \citep{ref3}. Another work \citep{MLRev2} compares various off-the-shelf machine learning tools for stock prediction. Further studies have proposed hybrid machine learning models, based on the use of multiple types of base classifiers that operate on a common input and a meta classifier that learns from base classifiers’ outputs to obtain a more precise stock return and risk predictions. In the work \citep{hybridFuzzyRev1}, the authors study the performance of financial technical indicators when used as inputs instead of machine learning based features for a neuro-fuzzy classifier. The study by \citep{DTWRev3} uses a sophisticated version of template matching called chart pattern recognition to identify profitable stocks. Strategies such as Bagging, Boosting and AdaBoost, can be also applied to create diversity in classifier combinations \citep{ref40,ref46}. For example, a hybrid weighted SVM and weighted KNN model for predicting stock market indices is proposed in \citep{ref16}. Similarly, a technique that combines Support Vector Regression (SVR), Random Forests and ANNs for predicting stock market index, is introduced in \citep{PATEL20152162}. 
%hybrid SOM (Self-Organizing Maps) using K means clustering for clustering the stocks whereas Support Vector Regression (SVR) is used to predict the future price and volatility for short trading cycles for better forecasts. 
Another study \citep{ref17} combines the statistical and probabilistic Bayesian Learning and the machine learning model ANN for the same. However, in all the aforementioned techniques, the relationship built between historical data and future value prediction may lack interpretation because of their ``black-box" property and, thus, the performance of these methods are directly related to the quality of the features. Moreover, with machine learning techniques, overfitting is a major issue owing to their capability of non-linear mapping and fitting.

%\noindent 
Deep learning based models have also been extensively used for solving stock forecasting problems. Recurrent Neural Networks (RNNs) are considered to be the most appropriate models for time-series analysis. LSTM is one such RNN which is regarded as the memory-mimicking model. Some studies use LSTM for the time-series stock forecasting \citep{ref18}. Another work uses LSTM on the technical indicators for the prediction \citep{ref19}. However, despite the great performance obtained, the time complexity of training RNN via backpropagation has encouraged the users for searching for more tractable models and solutions.
%\noindent 
CNNs constitute another important deep learning model, apart from RNNs, which have been used profusely and have performed well in the stock time-series forecasting, especially 2-D CNNs. In \citep{ref1}, the said techniques have been used on stock prices for forecasting. A slightly different input is used in \citep{ref20}, instead of using the standard variables (opening, closing, high, low and NAV), it uses high frequency data for forecasting major points of inflection in the financial market. In another work \citep{ref21}, a similar approach is used for modeling exchange traded fund (ETF). The 2-D CNN model performs similarly as LSTM or the standard multi-layer perceptron \citep{ref22,ref23} while being simpler to train. This apparent lack of performance improvement may be owing to the incorrect choice of CNN model, since these studies model an inherently 1D time series as an image. 

\subsection{Information Fusion}

 Many real world domains raise problems pertaining to the need for the fusion of information from multiple sources. Consider the problem of demand forecasting which requires estimating the power consumption at a future point given the available information until the current instant. At the building level forecasting, the inputs are usually power consumption, weather (temperature, humidity), and occupancy. This is a crucial problem in smart grids that ranges from planning electricity generation to preventing non-technical losses. Another area is biomedical signal analysis, for example, the problem of blood pressure estimation. The inputs are usually from two sources, namely the electrocardiogram (ECG) and pulsepleithismogram (PPG) \citep{ref33}, and the goal is to estimate the systolic and diastolic pressures. 
 %Another problem that requires inputs from ECG, PPG and skin moisture content is estimating stress and fatigue. 
 Transportation is also one such domain that needs the fusion of information from many sources to build intelligent transportation systems (ITS) \citep{transport,SAADI2018352}. This is needed to improve passenger safety, reduced transportation time and fuel consumption, etc. 
%  In the same domain, the work \citep{taxidemand} deals with the problem of forecasting the taxi demand in the event areas. It is done by fusing the publicly available data and time-series data using deep learning techniques. 

 Image fusion is another area where the information from two or more images of an object has to be integrated into a single image that is more informative and appropriate for visual perception or computer analysis. It finds great application in medical imaging. One can mention for instance the fusion of MRI (Magnetic Resonance Imaging) and PET (Positron Emission Tomography) images using IHS (Intensity Hue Saturation) and RIM (Retina-Inspired Models) fusion methods to improve the functional and spatial information content of the PET images \citep{ref34}. 
 
%  Another area that focuses on multi-channel data fusion is multi-sensor video fusion, again finding application in the medical domain. It uses the scanpath assessment for the visible and infrared side-by-side and fused video displays \citep{ref35}. 
 
 Deep learning has been widely used for analyzing multi-channel / multi-sensor signals. In such studies, all the sensors are stacked one after the other to form a matrix using 2-D CNN further to analyze these signals. For example, \citep{ref24} uses the same explained model to analyze human activity recognition from multiple body sensors. Note that it must be distinguished from the studies mentioned before \citep{ref1,ref20,ref21,ref22,ref23}, as the images in \citep{ref24} are not formed from stacking windowed signals from the same signal one after the other, they are formed by stacking signals from different sensors. Note, however, that \citep{ref24} does not account for any temporal modeling. This is rectified in \citep{ref7} where 2-D CNN is used on a time series window. The different windows are finally processed by GRU, thus explicitly incorporating time series modeling. In the aforesaid studies, there is however no explicit fusion framework. The information from raw signals is fused to form matrices and treated by 2-D convolutions. A true fusion framework was proposed in \citep{ref26}. Here, the fusion was happening at the feature level and not in the raw signal level as was in \citep{ref24,ref7}. 
 
 Multi-modal data processing is another area that makes use of deep learning based fusion techniques. Although this problem is not multi-channel data processing \emph{per se}, we will briefly review here some studies on this topic. In \citep{ref27} a fusion scheme is proposed for audio-visual analysis, that uses a fusion scheme for deep belief network (DBN) and stacked autoencoder (SAE) for fusing the audio and video channels. Each of the said channels is processed separately and connected by a fully connected layer to produce fused features. These fused features are further processed for inference. %The multi-modal deep network has also been utilized in script identification task at the word level \citep{ref28}. 
 %The multi-modal deep network has also been utilized in script identification task at the word level. The network takes both offline and online modality of the data as input to explore the information from both the modalities jointly for the script identification task \citep{ref28}. 
 The problem of video based action recognition is addressed in \citep{ref29}. It does not require audio data for the task; rather it proposes a fusion scheme for incorporating temporal information (processed by CNN) and spatial information (also processed by CNN). Experiments were carried out with different levels of early and late fusion. The fusion of multi-channel image dataset has also been investigated. In \citep{ref30}, a fusion scheme is proposed for processing color and depth information (via 3-D and 2-D convolutions, respectively) with the objective of action recognition. In \citep{ref31}, the authors consider fusing hyperspectral data (high spatial resolution) with Lidar (depth information), with the consequence of better classification results. In \citep{ref32}, it was shown that by fusing deeply learnt features (from CNN) with handcrafted features via a fully connected layer, can improve analysis tasks.

It is worthy to point out that the aforementioned time-series data based fusion studies do not process the time-series data as 1-D but as 2-D image/matrix. In the context of financial time-series, the state-of-the-art methods seem mostly based either on statistical and machine learning models or CNNs. For the former, the relationship built between historical data and future value prediction may lack interpretation because of their ``black-box" property; and hence, the performance of the methods is directly related to the quality of the features. While in the latter case of CNNs, there is no guarantee of unique filters learnt. In this work, we propose a novel framework that can tackle those issues.

\section{Proposed Technique}

This paper introduces a novel supervised framework for multi-channel data representation learning. A crucial element of the latter is our recently introduced CTL \citep{ref36}. For clarity, we first recall the important steps of the CTL technique. Then, we propose an extension of this approach in order to handle a multi-layer architecture. Finally, we present the overall SuperDeConFuse~(SDCF) architecture. 

\subsection{Convolutional Transform Learning}
As introduced in our seminal paper \citep{ref36}, CTL learns some filters $\left(t_m\right)_{1 \leq m \leq M}$ operated on samples $\left(s^{(k)}\right)_{1 \leq k \leq K}$ to generate the features $\big(x_m^{(k)}\big)_{1 \leq m \leq M,1 \leq k \leq K}$. The inherent learning model is expressed by convolution operations (assuming suitable padding) defined as
\begin{equation}\label{eq1}
(\forall m \in \{ 1,\ldots,M\}\;, \forall k \in \{ 1, \ldots, K\})\qquad {t_m} * {s^{(k)}} = x_m^{(k)}.
\end{equation}

A regularization is imposed on the filters to improve the representation ability and limit the overfitting issues, following from the original study on transform learning \citep{ref44}. Also, non-negativity constraint is imposed on the features, as it is commonly done in CNNs. The convolutional filters and the representation coefficients are learnt from the data during training. This is expressed as the following optimization problem:
\begin{multline}
\label{eq2}
\minimize{{(t_m)_m},(x_m^{(k)})_{m,k}} \frac{1}{2}\sum\limits_{k = 1}^K {\sum_{m = 1}^M \Big( {\big\| {{t_m} * {s^{(k)}} - x_m^{(k)}} \big\|_2^2} }  + \psi( x_m^{(k)} ) \Big)
\\
+ \mu \sum_{m = 1}^M \left\| t_m \right\|_2^2 - \lambda \log \det \big( [ {{t_1}\;\ldots\;{t_M}} ] \big),
\end{multline}
where $\psi$ is a suitable penalization function, and $(\mu,\lambda)$ are positive hyperparameters. %, and $\iota_+$ denotes the indicator function of the non-negative orthant, equals to $0$ if its input has non-negative entries, and $+\infty$ otherwise. 
It should be noted that the regularization term promotes unique filters to be learnt, something that is not easy to guarantee in CNNs. We can rewrite equivalently Equation \eqref{eq2} in matrix notation as\footnote{Note that $T$ is not necessarily a square matrix. By an abuse of notation, we define the ``log-det" of a rectangular matrix as the sum of logarithms of its singular values, taking infinity value as soon as one of those is non positive.}
\begin{equation}
\label{eq:onelayer}
F(T,X) = \frac{1}{2}\left\| T \star S - X \right\|_F^2 + \Psi(X) 
%+ {\iota _+}\left( X \right)  
+ \mu \left\| T \right\|_F^2 - \lambda \log \det \left( T \right),
\end{equation}
where $T=\begin{bmatrix}t_1 & \dots & t_M\end{bmatrix}$, $S=\begin{bmatrix}s^{(1)} & \dots& s^{(K)}\end{bmatrix}^\top$, $X=\begin{bmatrix} x_1^{(k)} & \dots & x_M^{(k)} \end{bmatrix}_{1\le k\le K}$, 
\begin{equation}
T \star S
=
\begin{bmatrix}
t_1 * s^{(1)}  & \dots & t_M * s^{(1)}\\
\vdots & \ddots&\vdots\\
t_1 * s^{(K)}& \dots & t_M * s^{(K)},
\end{bmatrix}
\end{equation}
and $\Psi$ amounts to applying the penalty term $\psi$ column-wise on $X$ and summing.

%The cost function in Problem \eqref{eq2} can be compactly rewritten as
% \begin{equation}
% \label{eq:onelayer}
% F(T,X) = \frac{1}{2}\left\| T*S - X \right\|_F^2 + \Psi(X) 
% %+ {\iota _+}\left( X \right)  
% + \mu \left\| T \right\|_F^2 - \lambda \log \det \left( T \right),
% \end{equation}
%where

A local minimizer to \eqref{eq:onelayer} can be reached efficiently using the alternating proximal algorithm \citep{ref41,ref42,ref43}, which alternates between proximal updates on variables $T$ and $X$. 
% For more details on the minimizer and proximal updates, the readers can refer to our paper \citep{deconfuse}.
The proximity operator \citep{ref45} at $\tilde x \in \mathcal{H}$, with $(\mathcal{H},\|\cdot\|)$ a Hilbert space, of a proper lower-semi-continuous convex function $\varphi : \mathcal{H} \to ] - \infty , + \infty ]$ is defined as
\begin{equation}
\label{eq4}
\operatorname{prox}_{\varphi}(\tilde x) = \mathop {\arg \min }\limits_{x \in \mathcal{H}} \varphi (x) + \frac{1}{2}\left\| {x - \tilde x} \right\|^2.
\end{equation}
Then, the alternating proximal algorithm for CTL reads:
\begin{equation}
\begin{array}{l}
{\rm{For\ }}{n} = 0,1,...\\
\left\lfloor \begin{array}{rl}
T^{[n + 1]} &= \operatorname{prox}_{\gamma_1 F(\cdot,X^{[n]})} \left(T^{[n]}\right)\\
X^{[n + 1]} &= \operatorname{prox}_{\gamma _2 F(T^{[n + 1]},\cdot)}\left( X^{[n]} \right)
\end{array} \right.
\end{array}
\end{equation}
with initializations $T^{[0]}$, $X^{[0]}$ of suitable dimensions, and ${\gamma _1}, {\gamma _2}$ some positive constants. For more details on the derivations and the convergence guarantees, the readers can refer to \citep{ref36}.

\subsection{Deep Convolutional Transform Learning}
\begin{figure}%[!htb]
\includegraphics[width=\textwidth]{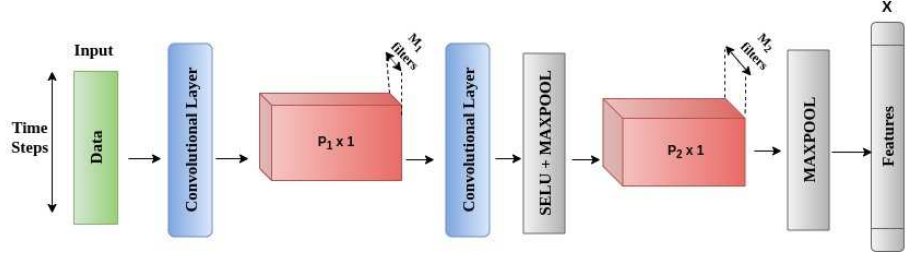}
\caption{Deep CTL architecture for $L=2$ layers.}
\label{sdctl}
\end{figure}

Deep CTL consists in stacking multiple convolutional layers on top of each other to generate the features, as shown in Figure \ref{sdctl}. Deep CTL depends on the key property that the solution $\widehat{X}$ to the CTL problem, assuming fixed filters $T$, can be reformulated as the simple application of an element-wise activation function. That is:
% To learn all the variables in an end-to-end fashion, deep CTL relies on the key property that the solution $\widehat{X}$ to the CTL problem, assuming fixed filters $T$, can be reformulated as the simple application of an element-wise activation function, that is 
\begin{equation}\label{eq:relu}
\operatorname*{argmin}_X F(T,X) = \Phi\big(T \star S\big),
\end{equation}
with $\Phi$ the proximity operator of $\Psi$ \citep{Combettes}. It is interesting to remark that, if $\Psi$ is the indicator function of the positive orthant, then $\Phi$ identifies with the famous rectified linear unit (ReLU) activation function. Many other examples of mapping between proximity operators and activation functions are provided in \citep{Combettes}. Consequently, we propose to compute deep features by stacking many such layers:
\begin{equation}\label{eq:deep_feats}
(\forall \ell\in\{1,\dots,L-1\})\qquad X_\ell = \Phi_\ell(T_\ell \star X_{\ell-1}),
\end{equation}
where we set $X_0 = S$.
%, and $\phi_\ell$ denotes a generic nonlinear activation followed by max-pooling. Note that we dropped the sparsity promoting regularization on the coefficients by setting $\beta=0$. In deep learning, the dimensionality of the coefficients usually reduces in each layer, and thus the representation is naturally compact. 
%Putting all together, deep CTL amounts to
Deep CTL consists of solving the problem%becomes while using all the explained elements together
\begin{align}\label{eq:dctl}
\minimize{T_1,\dots,T_L,X} F_{\rm conv}(T_1,\dots,T_L,X\,|\,S)
\end{align}
with
\begin{align}
F_{\rm conv}(T_1,\dots,T_L,X\,|\,S) &= \frac{1}{2} \| T_L \star \Phi_{L-1}(T_{L-1} \star \dots \Phi_1(T_1 \star S)) - X\|_F^2 \nonumber\\
&+ \Psi(X) + \sum_{\ell=1}^{L}\big(\mu||T_\ell||^2_F - \lambda\log\det(T_\ell)\big).
\end{align}
Deep CTL can thus be viewed as a natural and simple extension of the one-layer CTL formulation in \eqref{eq:onelayer}. 

\subsection{Our Proposed Approach - SuperDeConFuse}

\begin{figure}[!htb]
\includegraphics[width=6.0in, height= 3.2in]%[width=5.7in,height=3.3in][scale=0.75,page=1]
{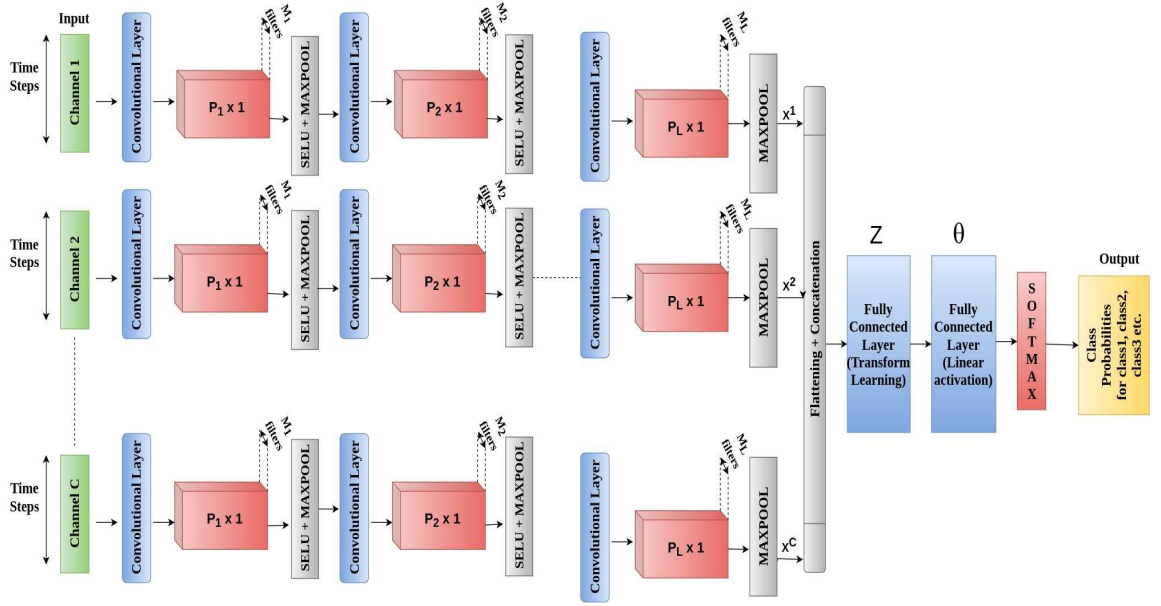}
\caption{SuperDeConfuse Architecture. The architecture is tested for $L=1,2,3,4$ layers. Here $P_1 \times 1,\ldots,P_L \times 1$ represents the kernel size used in each layer $\ell \in \{1,\ldots,L\}$. Here, maxpooling is not performed after layer $4$ due to the small window size/input sequence length.}
\label{sudctl}
\end{figure}

We now present our novel approach, \emph{SuperDeConFuse} (SDCF), which is a supervised fusion framework for multi-channel time-series stock data. This framework takes the channels of input data samples to separate branches of convolutional layers, leading to multiple sets of channel-wise features. The features obtained are thus decoupled. In order to couple (i.e., fuse) them, these are concatenated and passed to a fully-connected layer, which yields a set of unique coupled features via transform learning. These features are then fed to another linear fully-connected layer. This provides features that are finally inputted to the softmax layer that yields the probabilities for the classes. The complete architecture, called SuperDeConFuse (SDCF), is shown in Figure \ref{sudctl}.

As the data considered is multi-channel, we learn a different set of convolutional filters $T^{(c)}_1,\ldots,T^{(c)}_L$ and features $X^{(c)}$ for each channel $c \in \{1,\ldots,C\}$. We also learn the (not convolutional) linear transform $\widetilde{T}=(\widetilde{T_c})_{1 \le c \le C}$ to fuse the channel-wise features $X=(X^{(c)})_{1\le c \le C}$, along with the corresponding fused features $Z$ at the same time. The latter task is carried out by the cost function 
\begin{equation}\label{eq:fusion}
F_{\rm fusion}(\widetilde{T}, Z, X) =\frac{1}{2}\Big\|Z-\sum^{C}_{c=1}{\rm flat}(X^{(c)}) \widetilde{T}_c \Big\|^2_F + \Psi(Z) + \sum_{c=1}^C\Big(\mu\Big\|\widetilde{T_c}\Big\|^2_F - \lambda\log \det(\widetilde{T_c}) \Big)
\end{equation}
where the operator ``$\operatorname{flat}$" transforms $X^{(c)}$ into a matrix where each row contains the ``flattened" features of a sample.

%Analyzing the same dimensionally, it means that if $D_\ell$ = output sample size at layer $\ell$ then $I = D_L M_L$ is the number of output features per sample at the last convolution layer $L$ which is the dimension for X then $O=\alpha I C$ (with $\alpha \in]0,1]$) is the number of output features per sample at the fully-connected layer ie. $\widetilde{T}_c \in\mathbb{R}^{I \times O}$ and $Z \in \mathbb{R}^{I \times O}$ per sample.

Further, we learn the weight matrix $\theta$ of a multiclass classifier which takes the input features $Z$ and yields the class probabilities. 
%linear fully connected layer features $\theta$ which has input as features $Z$ (learnt at previous fully connected layer) and and the product of $Z$ and $\theta$ gives the scores / logits. We pass these scores / logits to the softmax layer yielding class-probabilities. 
The cross-entropy (CE) loss associated with the final classification is given by 
\begin{equation}\label{eq:crossEntropy1}
F_{\textrm{CE}}(\theta,Z \,|\, y) = \sum_{k=1}^{K}\log\Big(\sum_{v=1}^V e^{z_k^\top (\theta_{v} - \theta_{y_k})}\Big),
\end{equation}
where $V$ is the number of classes, $\theta_v$ is the $v$-th column of matrix $\theta$, $z_k^\top$ is the $k$-th row of matrix $Z$, and $y_k \in \{1,\dots,V\}$ is the label of the $k$-th sample.
This finally leads to the joint optimization problem defined as
\begin{equation}\label{eq:joint}
\minimize{(T,X,\widetilde{T},Z,\theta)} \underbrace{\sum_{c=1}^{C}F_{\rm conv}(T_1^{(c)},\ldots,T_L^{(c)}, X^{(c)} | S^{(c)}) + F_{\rm fusion}(\widetilde{T},Z,X) + F_{\rm CE}(\theta,Z\,|\, y).}_{J(T, X, \widetilde{T}, Z, \theta)}
\end{equation}

Conclusively, our formulation aims at jointly training the channel-wise convolutional filters $T_l^{(c)}$, the fusion coefficients $\widetilde{T}$, and the multiclass classifier $\theta$ in an end-to-end fashion. We explicitly learn the features $X$ and $Z$ subject to the regularization $\Psi$, so as to avoid the problem of dead neurons. Moreover, the ``log-det" regularization on both $T_l^{(c)}$ and $\widetilde{T}$ breaks the symmetry and enforces the diversity in the learnt transforms, whereas the Frobenius regularization keeps the transform coefficients bounded.

% \begin{equation}\label{eq:crossEntropy}
% F_{\textrm{labels}} = -\sum_{k=1}^{K}\sum_{i=1}^{I}y_{k,i}\log(p_{k,i})	
% \end{equation}
% where $I = 3$ classes (BUY, HOLD, SELL), $y_{k,i}$ is binary indicator ($0$ or $1$) if class label k is the correct classification for observation k, $p_{k,i}$ is the predicted probability observation k is of class i given by the final layer with the Softmax function as-
% \begin{equation}\label{eq:softmax}
% p_{k,i} = \frac{\exp{(g_{k,i}(\theta))}}{\sum_{j=1}^{N}\exp{(g_{k,j}(\theta))}}
% \end{equation}
% and $g_{k,i}(\theta) = \theta^{T}Z + b$. Here $\theta$ is the weights learnt from a linear fully connected layer. %and then finally a softmax layer. 
% This finally leads to the joint optimization problem :
% \begin{equation}\label{eq:joint}
% \minimize{(T,X,\widetilde{T},Z)} \underbrace{\sum_{c=1}^{C}F_{\rm conv}(T_1^{(c)},\ldots,T_L^{(c)}, X^{(c)} | S^{(c)}) + F_{\rm fusion}(\widetilde{T},Z,X) + F_{\rm labels}()}_{J(T, X, \widetilde{T}, Z)}
% \end{equation}

\subsection{Optimization algorithm}\label{sec:optimization}
%Let us now present the solution of the Problem \eqref{eq:joint}. 
We propose to find a local minimizer to the nonconvex Problem \eqref{eq:joint} through the projected (sub)gradient descent, whose iterations read:
% \begin{equation}
% \begin{array}{l}
% {\rm{For\ }}{n} = 0,1,...\\
% \;\left\lfloor \begin{array}{rl}
% T^{[n + 1]} &= T^{[n]} - \gamma \nabla_T J(T^{[n]}, X^{[n]}, \widetilde{T}^{[n]}, Z^{[n]})\\
% X^{[n + 1]} &= \mathcal{P}_+\big(X^{[n]} - \gamma\nabla_X J(T^{[n]}, X^{[n]}, \widetilde{T}^{[n]}, Z^{[n]})\big)\\
% \widetilde{T}^{[n + 1]} &= \widetilde{T}^{[n]} - \gamma\nabla_{\widetilde{T}} J(T^{[n]}, X^{[n]}, \widetilde{T}^{[n]}, Z^{[n]})\\
% Z^{[n + 1]} &= \mathcal{P}_+\big( Z^{[n]} - \gamma\nabla_Z J(T^{[n]}, X^{[n]}, \widetilde{T}^{[n]}, Z^{[n]})\big)\\
% \end{array} \right.
% \end{array}
% \end{equation}
% \noindent with initialization $T^{[0]}, X^{[0]}, \widetilde{T}^{[0]}, Z^{[0]}$,  $\gamma>0$, and $\mathcal{P}_+ = \max\{\cdot,0\}$. 

\begin{equation}
\begin{array}{l}
{\rm{For\ }}{n} = 0,1,...\\
\;\left\lfloor \begin{array}{rl}
T^{[n + 1]} &= T^{[n]} - \gamma \nabla_T J(T^{[n]}, X^{[n]}, \widetilde{T}^{[n]}, Z^{[n]},\theta^{[n]})\\
X^{[n + 1]} &= \mathcal{P}_+\big(X^{[n]} - \gamma\nabla_X J(T^{[n]}, X^{[n]}, \widetilde{T}^{[n]}, Z^{[n]},\theta^{[n]})\big)\\
\widetilde{T}^{[n + 1]} &= \widetilde{T}^{[n]} - \gamma\nabla_{\widetilde{T}} J(T^{[n]}, X^{[n]}, \widetilde{T}^{[n]}, Z^{[n]}, \theta^{[n]})\\
Z^{[n + 1]} &= \mathcal{P}_+\big( Z^{[n]} - \gamma\nabla_Z J(T^{[n]}, X^{[n]}, \widetilde{T}^{[n]}, Z^{[n]},\theta^{[n]})\big)\\
\theta^{[n + 1]} &= \theta^{[n]} - \gamma\nabla_{\theta} J(T^{[n]}, X^{[n]}, \widetilde{T}^{[n]}, Z^{[n]},\theta^{[n]})\\
\end{array} \right.
\end{array}
\end{equation}
with $\mathcal{P}_+ = \max\{\cdot,0\}$ (applied element-wise). In practice, we initialize it with some random matrices $T^{[0]}, X^{[0]}, \widetilde{T}^{[0]}, Z^{[0]}, \theta^{[0]}$, we choice a suitable stepsize $\gamma>0$, and we evaluate numerically the gradient step with the accelerated scheme initially introduced for the ADAM method in \citep{adam}.

There are two remarkable advantages of the proposed optimization approach. Firstly, we depend on automatic differentiation \citep{ref38} and stochastic gradient approximations to efficiently solve Problem \eqref{eq:joint}. Secondly, any sub-differentiable activation function $\Phi$ in \eqref{eq:relu} can be plugged into our model, for instance SELU \citep{NIPS2017_6698} or Leaky ReLU \citep{Maas}. This flexibility will play a key role in the performance, as shown in the experimental section.  

\subsection{Computational Complexity of Proposed Framework - SuperDeConFuse(SDCF)}
Table \ref{tab:complexity} summarizes the computational complexity of SuperDeconFuse(SDCF) architecture, both for training and test phases. We report the cost incurred for one input sample, either at an iteration of the training algorithm or at the testing phase. It is to be noted that the computational complexity of SDCF architecture is comparable to that of a standard CNN. The log-det regularization is the only addition that requires to compute the truncated singular value decomposition of $T_\ell^{(c)}$ and $\widetilde{T}_c$. However, as the size of these matrices is determined by the filter size, the number of filters, and the number of output features per sample, the training complexity is not worse than that of a CNN.

\begin{table}[!h]
    \caption{\textbf{Time complexity in training and test phases (for one input sample)}}
    \label{tab:complexity}
    \centering\small
    \begin{tabular}{|c|l|c|l|}
    \hline
         \textbf{Phase} &
         \textbf{Steps} & 
         \textbf{Time}&
         \textbf{Dimension}
         \\
         & & \textbf{Complexity} & \textbf{Description}
         \\
         \hline
         \textbf{Training}
         &
         1. Convolution layers
         &
         $\mathcal{O}(P_\ell D_\ell M_\ell C)$ 
         &
         %{\scriptsize $t_m^{(c)} \in \mathbb{R}^P$}
         \\
         %\textbf{phase} 
         & 
         2. Fully-connected (f.-c.) layer
         & 
         $\mathcal{O}(I^2 C^2)$ 
         &
         {\scriptsize $S^{(c)} \in \mathbb{R}^{K\times D}$}
         \\
         & 3. Frobenius norm on conv.\ layers & $\mathcal{O}\big(P_\ell M_\ell C\big)$
         & {\scriptsize $T_\ell^{(c)}\in\mathbb{R}^{P_\ell\times M_\ell}$}
         \\
         & 4. Frobenius norm on f.-c.\ layer 
         & $\mathcal{O}(I^2 C^2)$ 
         & {\scriptsize ${\rm flat}(X^{(c)})\in\mathbb{R}^{K\times I}$}
         \\
         & 5. log-det on conv.\ layers
         & $\mathcal{O}(P_\ell^2 M_\ell C)$ %$\mathcal{O}(\min\{P_\ell^2M_\ell C, P_\ell M_\ell^2 C\})$ 
         & {\scriptsize $\widetilde{T_c} \in\mathbb{R}^{I \times O}$}
         \\
         & 6. log-det on f.-c.\ layer & $\mathcal{O}(I^3C^2)$ & {\scriptsize $Z \in\mathbb{R}^{K \times O}$}
         \\
         & 7. output layer (classifier) & $\mathcal{O}(V)$ & {\scriptsize $\theta \in\mathbb{R}^{O \times V}$}
         \\
         \hline 
         \textbf{Test}
         & 
         \begin{tabular}[h]{l}
         Step 1 + Step 2 + Step 7. 
         \end{tabular}
         &
         See above.
         &
         \\ 
    \hline
    \end{tabular}
    \begin{tablenotes}
        \item {\scriptsize $D$ = input sample size -- $K$ = num.\ of samples -- $C$ = num.\ of channels -- $L$ = num.\ of layers}
        \item{\scriptsize $P_\ell$ = filter size at layer $\ell$ -- $M_\ell$ = num.\ of filters at layer $\ell$ -- $D_\ell$ = output sample size at layer $\ell$}
        \item {\scriptsize $I = D_L M_L$ is the num.\ of output features per sample and per channel at last convolution layer}
        \item {\scriptsize $O=\alpha IC$ (with $\alpha \in]0,1]$) is the num.\ of output features per sample at the fully-connected layer}
        \item {\scriptsize $V$ = num.\ of classes}
    \end{tablenotes}
\end{table}

\section{Methodology}
\subsection{Dataset Description}
%\noindent 
The dataset consists of 15 Indian stocks that fall under the National Stock Exchange (NSE) and the Bombay Stock Exchange (BSE). The stock symbols end with .NS if fall under NSE and .BO for BSE otherwise. These stock symbols are taken from Yahoo finance symbols data available publicly. The data is made of day-wise readings for the past 22 years i.e. from 1998 - 2019 is collected using the in-built python module web and the Yahoo API end-point internally. At the time of data collection, the data for the year 2019 was not a complete year’s data, so that there were some missing values for some raw features. Thus, we have not used the data for 2019 in our experiments for the sake of simplicity. The dataset includes stocks from multiple sectors such as Indian consumer products manufacturers (e.g., HINDUNILVR.NS), oil and gas (e.g. CAIRN.NS), pharmaceuticals (e.g. AUROPHARMA.NS, DRREDDY.NS), mining and metal industry (e.g. NATIONALUM.BO). 

\subsection{Labeling}
After curating the dataset for 15 stocks with values for the features - date, symbol, adjusted (adj.)\ close price, opening price, low price, high price, and net asset value, we have labeled the data. We will call the adj.\ close price as Close Price in the rest of the paper. In the labeling phase, we manually assign the labels to the daily close prices as Buy (0), Hold (1), Sell (2). The labels are determined by performing a grid search on the list of holding percentages to identify the percentage change for which the stocks should be held to maximize the annualized returns for the company. Algorithm \ref{alg:algorithm} gives the details of the labeling process.

\begin{algorithm}[H]
\caption{Labelling Method}
\label{alg:algorithm}
\textbf{Input} : CP - Array of Closing Prices, S - stock/symbol \\
\textbf{Parameter} : X - array of K holding percentages, \\ \hspace{1cm} NUMDAYS - number of days for the current symbol or len(CP)\\
Labels - 2D array of size K x NUMDAYS\\
\textbf{Output} : FinalLabels - Labelled Dataset for S
 %[1] enables line numbers
\begin{algorithmic}[1]
\STATE AR = [ ] //it is of size K\\
%\STATE NUMDAYS = len of days in current symbol\\
\FOR {$k = 0,1,2,\ldots,K-1$}
\FOR {$n = 0,\ldots,NUMDAYS-1$}
\STATE change = abs$(({CP[n+1]-CP[n]}/CP[n])*100)$ //where CP[n+1] is the next day closing price\\
\IF {change $>$ X[k]}
\IF {$CP[n+1] > CP[n]$}
\STATE \textbf{label} == ``Sell"
\ELSE
\STATE \textbf{label} == ``Buy"
\ENDIF
\ELSE
\STATE \textbf{label} == ``Hold"
\ENDIF
\STATE Labels[k].append(label)
\ENDFOR
\STATE ar = AnnualisedReturn(Labels[k],CP)\\
\STATE AR.append(ar)
\ENDFOR
\STATE maxAr = Max(AR), maxIndex = index(Max(AR))\\
\STATE HoldPercentage = X[maxIndex] \\
\STATE FinalLabels = Labels[maxIndex] \\
\STATE \textbf{return} FinalLabels
\STATE Repeat all steps till 22 for all the Stocks/Symbols in the dataset.
% \STATE \textbf{return} FinalLabels
\end{algorithmic}
\end{algorithm}

\subsection{Training Details}
\begin{figure}[!htb]
\includegraphics[width=\textwidth,height=2.3in]{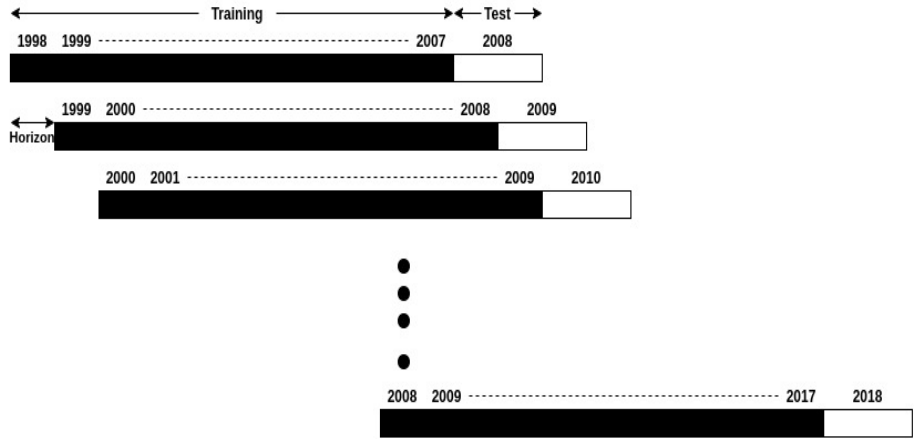}
\caption{Sliding walk-forward validation technique used for hyperparameters tuning}
\label{swfv}
\end{figure}

%\noindent 
We use the sliding walk forward validation technique which is used as the cross-validation technique in case of time-series data also shown in Figure \ref{swfv}. As can be seen from Figure \ref{swfv}, we use 10 years of data for training and the subsequent 1 year data for testing, i.e., the stock data from 1998-2007 is for training and the year 2008 for testing. Then we slide the training window by 1 year which implies that we next train it from 1999-2008 and test it on the following year 2009 data and this period is called as horizon. In general, we train for 10 years, test it for the following year and then slide it by a 1 year horizon and again train and test and so on till the year 2018. Thus, 11 years of data from 2008 - 2018 are used as test data. This way, we have 11 models and we select the set of hyperparameters that give the best results across all the 11 models. The set of hyperparameters that we tune includes $\mu,\lambda$, kernel sizes, number of filters/kernels, learning rate, weight decay of the Adam optimizer, batch size, and number of epochs. Additionally, we randomly initialize the weights for each stock’s training. This appears here as a very efficient technique to analyse the robustness of the architecture. In other words, we calculate the model performance every time a year's data becomes available for testing and we use previous 1 year test data for training. We standardize the training and the test data using Normalizer from Python library as prices and the NAV features/channels have a varied range of values.

\section{Experimental Evaluation }
%\noindent 
We carry out experiments on the real world problem of stock trading. Stock trading is a classification problem, where the decision whether to buy or hold or sell a stock has to be taken at each time. The problem makes a decision that if the price of a stock at a later date is expected to increase, the stock must be bought; and if the stock price is expected to go down, the stock must be sold; and if there is no change in the price then it should be held, i.e., do nothing until the price increases. This is done in a way so as to maximize the annualized returns from the stock for the company's profit as mentioned in the labeling process.

%\noindent 
We use the five raw inputs for both the tasks, namely open price, close price, high, low and net asset value (NAV). We chose to stay with the raw values. However, one could compute technical indicators based on the raw inputs \citep{ref1} but raw values allow here to keep up with the essence of the true nature of representation learning.
% To keep up with the true nature of representation learning, we chose to stay with the raw values rather than the 
% One could compute technical indicators based on the raw inputs \citep{ref1}. But, in keeping with the essence of true representation learning, we chose to stay with those raw values. 
Each of the five inputs is processed by a separate 1D processing pipeline. Each of these pipelines produces a flattened output (Figure \ref{sudctl}). These flattened outputs are then concatenated and fed for fusion into the Transform Learning layer acting as the fully connected layer (Figure \ref{sudctl}). Further, this is connected to another linear fully connected layer and finally, there is a softmax function. The softmax function gives the classification output which consists of the class probabilities for the three classes (BUY, HOLD and SELL). 
% The softmax layer/node is the classification node which yields the class probabilities for the three classes (BUY, HOLD and SELL). 
%\\
%\noindent 

We extend the architecture by adding CTL layers to 4 layers deep SDCF architectures. The details for all the four architectures are briefed in Table \ref{architecture}. Maxpooling halves the input sequence length/window size/Time Steps with its every operation. Thus, after 3 layers, the size is reduced to the value
%an extent
that it cannot be employed after the $4^{th}$ CTL layer; and, hence, the architecture with 4 CTL layers of SDCF will not have maxpooling operation after layer 4. This is due to the small window size. Also, for making predictions on any day, the past 10 days will be analysed through the model which are labeled as Time Steps shown in Figure \ref{sudctl}. %We call this as \emph{window size} too. 
Additionally, to avoid the data leak, we do not predict the stock trading signal for the first 10 days of every test year. The predictions from every year totaling to 11 years will be saved and further, the metrics will be computed to analyse the performance of our model. We will compute two sets of metrics here, namely (i) classification metrics and (ii) financial metrics.
\begin{enumerate}%[label=(\alph*)]
\item[(i)] Classification Metrics -
This set of metrics includes class-wise F1 score, Precision and Recall to assess the performance from a classification point of view. We also calculate the weighted F1 Score, Precision and Recall to account for the class imbalance for every stock. Note that, in such case, the F1 score is not equivalent to the harmonic mean of Precision and Recall since it is weighted. 

\item[(ii)] Financial Metrics -
We also evaluate the performance of our framework and state-of-the-art from the financial point of view. We calculate, in specific, the Annualized Returns(AR) which is calculated using the predictions from all the models. The AR value will be calculated as mentioned in \citep{ref1}. The starting capital will be Rs 10,00,00,000.0 and transaction charges will be Rs 10. We will use Indian currency to calculate the AR values since we have used all the Indian stocks. Note, however, that our metric is versatile and could be used to evaluate the model in any currency depending on the stocks analysed. 
\end{enumerate}

%Table 1 : Architecture details
\begin{ThreePartTable}
\footnotesize
\begin{longtable}{|c|p{5.5cm}|l|}
\caption{\textbf{Hyperparameters for the different instances of the proposed SDFC network (see Figure \ref{sudctl} for the general overview) used in the experimental section.}}\label{architecture}\\
%\begin{tabular}{|c|p{6.9cm}|l|}
\hline
\textbf{Method} &  \textbf{Architecture Description} & \textbf{Other Parameters}\\
% \hline
\hhline{|=|=|=|}
\begin{tabular}[h]{c}
SDCF 1L
% SDCF \\
% 1 Layer 
\end{tabular}
& 
\begin{tabular}[h]{l}
\(
5 \times \begin{cases}
\textbf{layer1} : \textbf{1D Conv} (1,16,3,1,1)\tnote{1}\\
         \textbf{Maxpool} (2,2)\tnote{2}\\
\end{cases}\)
% \newline
% \newline
\\
\textbf{layer2} : \textbf{Fully Connected (TL)\tnote{3}}
            %     \\
            %   \qquad \qquad \textbf{(Transform Learning)}
% \newline
\\
\textbf{layer3} : \textbf{Fully Connected (Linear)}
% \newline
\\
\textbf{Softmax}%\\%\newline
\end{tabular}
& 
\begin{tabular}[h]{l}
$Learning Rate = 0.001$,\\%\newline
% epochs = 100, \\%\newline 
$\lambda = 0.01, \mu = 0.0001$ \\%s\newline
$epochs = 100$, \\
\textbf{Optimizer Used}: \textbf{Adam}\\
**with parameters**\\%\newline
$(\beta1,\beta2) = (0.9,0.999)$, \\%\newline 
weight\_decay = 1e-4, \\%\newline 
epsilon = 1e-8  
\end{tabular}
\\
%\hline
% \cline{1-2}
% \noalign{\vskip\doublerulesep
%          \vskip-\arrayrulewidth}
% \cline{1-2}
% \cmidrule{1-2}\morecmidrules\cmidrule{1-2}
\hhline{|=|=|~|}
\begin{tabular}[h]{c}
% SDCF \\
% 2 Layers 
SDCF 2L
\end{tabular}
& 
\begin{tabular}[h]{l}
\(
5 \times \begin{cases}
\textbf{layer1} : \textbf{1D Conv} (1,8,3,1,1)\tnote{1}\\
         \textbf{SELU} + \textbf{Maxpool} (2,2)\tnote{2}\\
\textbf{layer2} : \textbf{1D Conv} (8,16,3,1,1)\tnote{1}\\
         \textbf{Maxpool} (2,2)\tnote{2}\\
\end{cases}\)
% \newline
% \newline
\\
\textbf{layer3} : \textbf{Fully Connected (TL)\tnote{3}} 
            %\\
             % \qquad \qquad \textbf{(Transform Learning)}
% \newline
\\
\textbf{layer4} : \textbf{Fully Connected (Linear)}
% \newline
\\
\textbf{Softmax}%\\%\newline
\end{tabular}
& 
% \begin{tabular}[h]{l}
% Hyperparameters same as \\
% for SDCF 1L. 
% Learning Rate = 0.001,\\%\newline
% % epochs = 100, \\%\newline 
% $\lambda = 0.01, \mu = 0.0001$ \\%s\newline
% $epochs = 100$, \\
% \textbf{Optimizer Used}: \textbf{Adam}\\
% **with parameters**\\%\newline
% $(\beta1,\beta2) = (0.9,0.999)$, \\%\newline 
% weight\_decay = 1e-4, \\%\newline 
% epsilon = 1e-8  
% \end{tabular}
\\
% \hline
% \cline{1-2}
\hhline{|=|=|~|}
\begin{tabular}[h]{c}
% SDCF \\
% 3 Layers
SDCF 3L
\end{tabular}
& 
\begin{tabular}[h]{l}
\(
5 \times \begin{cases}
\textbf{layer1} : \textbf{1D Conv} (1,4,11,1,5)\tnote{1}\\
         \textbf{SELU} + \textbf{Maxpool} (2,2)\tnote{2}\\
\textbf{layer2} : \textbf{1D Conv} (4,8,7,1,3)\tnote{1}\\
         \textbf{SELU} + \textbf{Maxpool} (2,2)\tnote{2}\\
\textbf{layer3} : \textbf{1D Conv} (8,16,3,1,1)\tnote{1}\\
         \textbf{Maxpool} (2,2)\tnote{2}\\         
\end{cases}\)
% \newline
% \newline
\\
\textbf{layer4} : \textbf{Fully Connected (TL)\tnote{3}}
            %     \\
            %   \qquad \qquad \textbf{(Transform Learning)}
% \newline
\\
\textbf{layer5} : \textbf{Fully Connected (Linear)}
% \newline
\\
\textbf{Softmax}%\\%\newline
\end{tabular}
& 
% \begin{tabular}[h]{l}
% Hyperparameters same as \\
% for SDCF 1L. 
% Learning Rate = 0.001,\\%\newline
% % epochs = 100, \\%\newline 
% $\lambda = 0.01, \mu = 0.0001$ \\%s\newline
% $epochs = 100$, \\
% \textbf{Optimizer Used}: \textbf{Adam}\\
% **with parameters**\\%\newline
% $(\beta1,\beta2) = (0.9,0.999)$, \\%\newline 
% weight\_decay = 1e-4, \\%\newline 
% epsilon = 1e-8  
% \end{tabular}
\\
% \hline
% \cline{1-2}
\hhline{|=|=|~|}
\begin{tabular}[h]{c}
% SDCF \\
% 4 Layers 
SDCF 4L
\end{tabular}
& 
\begin{tabular}[h]{l}
\(
5 \times \begin{cases}
\textbf{layer1} : \textbf{1D Conv} (1,4,13,1,6)\tnote{1}\\
         \textbf{SELU} + \textbf{Maxpool} (2,2)\tnote{2}\\
\textbf{layer2} : \textbf{1D Conv} (4,8,11,1,5)\tnote{1}\\
         \textbf{SELU} + \textbf{Maxpool} (2,2)\tnote{2}\\
\textbf{layer3} : \textbf{1D Conv} (8,16,9,1,4)\tnote{1}\\
         \textbf{SELU} + \textbf{Maxpool} (2,2)\tnote{2}\\  
\textbf{layer4} : \textbf{1D Conv} (16,32,5,1,2)\tnote{1}\\
\end{cases}\)
% \newline
% \newline
\\
\textbf{layer5} : \textbf{Fully Connected (TL)\tnote{3}}
            %     \\
            %   \qquad \qquad \textbf{(Transform Learning)}
% \newline
\\
\textbf{layer6} : \textbf{Fully Connected (Linear)}
% \newline
\\
\textbf{Softmax}%\\%\newline
\end{tabular}
& 
% \begin{tabular}[h]{l}
% Hyperparameters same as \\
% for SDCF 1L. 
% Learning Rate = 0.001,\\%\newline
% % epochs = 100, \\%\newline 
% $\lambda = 0.01, \mu = 0.0001$ \\%s\newline
% $epochs = 100$, \\
% \textbf{Optimizer Used}: \textbf{Adam}\\
% **with parameters**\\%\newline
% $(\beta1,\beta2) = (0.9,0.999)$, \\%\newline 
% weight\_decay = 1e-4, \\%\newline 
% epsilon = 1e-8  
% \end{tabular}
\\
\hline
%\end{tabular}
\end{longtable}
\begin{tablenotes}
\item[1] \small{(in_planes, out_planes, kernel\_size, stride, padding)}
\item[2] \small{(kernel\_size, stride)}
\item[3] \small{TL - Transform Learning}
\item \small{L - \#CTL layers}
\end{tablenotes}

\end{ThreePartTable}
%\end{table}

%\noindent 
We compare with three state-of-the-art time series based analysis models, out of which two techniques present the models proposed specifically for financial stock trading - CNN-TA \citep{ref1} and MFNN \citep{ref2}; and the last technique presents a generic model for time-series based data - FCN(Fully Convolutional Network) \citep{ref39}. The latter is used as it helps understand how generic the proposed model is if compared against both specific stock trading based and general time-series models. In all the techniques, processing pipelines are based on CNN. Other than CNN, MFNN \citep{ref2} is also based on the RNN type of network - LSTM. In \citep{ref1}, the data is not used raw but processed as technical indicator values and passed as an image, hence uses 2D CNN whereas, in FCN \citep{ref39}, the data is processed via 2D CNN. The same hyperparameters for the benchmark techniques are used as given in the study except for FCN which is best tuned for our data.
We have also compared our model to the simple CNN with the architecture same as that of our framework i.e. $3$ convolutional layers deep architecture and used the same hyperparameters too except the kernel sizes of $P_1 = 11$, $P_2 = 9$ and $P_3 = 7$ for the convolutional layers $\ell = 1, 2$ and $3$ (padding size is $P_{\ell}/2$). The difference lies in the objective function of the convolutional learning in both the techniques i.e. our $3$ layers deep SDCF and $3$ layers deep simple 1D CNN. This is done to analyze the performance difference between the two supervised learning techniques. Additionally, we chose the architecture for CNN having $3$ convolutional layers, since the results depleted after $3$ convolutional layers for our framework and were best with $3$ layers.   

\subsection{Classification Analysis}
\label{comp_analysis}
%\noindent 
As mentioned previously, we first look at the Classification performance of our models. We test the framework for shallow - 1 CTL layer and deeper versions - 2, 3 and 4 CTL layers. The generated features from the fully connected layers are passed to the softmax and we get the probabilities for all the classes. The one with the maximum probability is selected as the predicted label. The performance is calculated for every class. Specifically, metrics - F1 Score, Precision and Recall are calculated for BUY, HOLD and SELL classes. The results are detailed in Table \ref{a_detailed_buy}, \ref{a_detailed_hold}, \ref{a_detailed_sell} in \ref{appendix_a}.

%\noindent 
Certain results are highlighted in bold or red. The first set of results in bold are the ones where one or more techniques for each metric give the best/greater than or equal performance. Analysing it in detail, we find that there are 8 stocks for which our proposed model performs greater than or equal to when compared with benchmark techniques for F1 score in case of the BUY class. Following the same, we find that the SDCF gives greater than or equal to performance for 13 stocks for precision and 5 stocks for recall metrics under the BUY class. Similarly, 7 stocks for F1 score, 7 stocks for precision and 5 stocks for recall in case of HOLD class and 7 stocks for F1 score, 11 stocks for precision and 6 stocks for recall in case of SELL class. Since we analyze our performance difference to understand the technique that has better supervised learning, we specifically look at the performance with CNN. CNN gives greater than or equal to performance for 2 stocks for each metric under BUY class. Similarly, there are 6, 1 and 9 stocks for the HOLD class and 2 stocks each for the metrics F1 score, precision and recall under SELL class.    
%\\
%\noindent 

Additionally, the other set of results in red indicate the performance where one of our proposed model versions gives the similar/next best performance under 0.02 error difference - err\_dif (let’s say) after one of the benchmarks i.e. $0.0 < \text{err}\_{\text{dif}} \le 0.02$. Adhering to the same, we observed that for BUY class, there is 1 stock each for metrics F1 score, precision and recall respectively. Likewise, for the HOLD class, there are 7, 4 and 5 stocks for F1 score, precision and recall metrics respectively; and for SELL class, we have 1 stock each for F1 score and recall metrics. We haven't, although, highlighted the results for CNN when it gives similar/next best performance but we present the statistics for the same here. Analyzing for CNN, there are 2 and 3 stocks for F1 score and precision under HOLD class. Observing these statistics, they indicate that the performance with our model is better than CNN for all three BUY, HOLD and SELL classes. 
%Together with the numbers related to both greater than or equal and next best performance, our model is better than CNN for all three classes.

\begin{table}[H]
%\begin{threeparttable}
\caption{\textbf{Summary of BUY Class Classification Results for Stock Trading}}\label{a_summary_buy}
\centering
\begin{tabular}{|c|c|c|c|}
% \begin{tabular}{|p{2cm}|p{1.5cm}|p{1.5cm}|p{1.5cm}|p{1.5cm}|p{1.5cm}|p{1.5cm}|p{1.5cm}|p{1.5cm}|p{1.5cm}|}
\hline
\textbf{Method} &  \begin{tabular}[h]{c}\textbf{Avg. BUY}\\\textbf{F1 Score}\end{tabular} &  \begin{tabular}[h]{c}\textbf{Avg. BUY}\\\textbf{Precision}\end{tabular} &  \begin{tabular}[h]{c}\textbf{Avg. BUY}\\\textbf{Recall}\end{tabular} \\
\hline
\begin{tabular}{l}SDCF 1L\end{tabular} & 0.0645 & 0.2182 & 0.0475 \\
\hline
\begin{tabular}{l}SDCF 2L\end{tabular} & 0.0916 & 0.2356 & 0.0683\\
\hline
\begin{tabular}{l}SDCF 3L\end{tabular} & 0.1091 & 0.2205 & 0.0854 \\
\hline
\begin{tabular}{l}SDCF 4L\end{tabular} & \textbf{0.1566} & \textbf{0.3242} & 0.1355 \\
\hline 
CNN & 0.0688 &	0.1179 & 0.0551\\
\hline
FCN & 0.0758 & 0.1446 & 0.0617 \\
\hline
CNN-TA & 0.1205 & 0.1611 & 0.1263 \\
\hline
MFNN & 0.0881 & 0.1672 & \textbf{0.2401} \\
\hline
\end{tabular}
%\end{threeparttable}
\end{table}

\begin{table}[H]
%\begin{threeparttable}
\caption{\textbf{Summary of HOLD Class Classification Results for Stock Trading}}\label{a_summary_hold}
\centering
\begin{tabular}{|c|c|c|c|}
\hline
\textbf{Method} &  \begin{tabular}[h]{c}\textbf{Avg. HOLD}\\\textbf{F1 Score}\end{tabular} &  \begin{tabular}[h]{c}\textbf{Avg. HOLD}\\\textbf{Precision}\end{tabular} &  \begin{tabular}[h]{c}\textbf{Avg. HOLD}\\\textbf{Recall}\end{tabular} 
\\
\hline
\begin{tabular}{c}SDCF 1L\end{tabular} & \textbf{0.7983} & 0.7091 & \textbf{0.9446} \\
\hline
\begin{tabular}{c}SDCF 2L\end{tabular} & 0.7912 & 0.7113 & 0.9164 \\
\hline
\begin{tabular}{c}SDCF 3L\end{tabular}  & 0.7813 & 0.7113 & 0.8842\\
\hline
\begin{tabular}{c}SDCF 4L\end{tabular}  & 0.6684 & 0.5950 & 0.7960 \\
\hline 
CNN & 0.7909 & 0.7090 & 0.9239\\
\hline
FCN & 0.7825 & 0.7119 & 0.9051\\
\hline
CNN-TA & 0.7686 & \textbf{0.7142} & 0.8557\\
\hline
MFNN &  0.5161 & 0.6425 & 0.5718 \\
\hline
\end{tabular}
%\end{threeparttable}
\end{table}

\begin{table}[H]
%\begin{threeparttable}
\caption{\textbf{Summary of SELL Class Classification Results for Stock Trading}}\label{a_summary_sell}
\centering
\begin{tabular}{|c|c|c|c|}
\hline
\textbf{Method} &  \begin{tabular}[h]{c}\textbf{Avg. SELL} \\\textbf{F1 Score}\end{tabular} &  \begin{tabular}[h]{c}\textbf{Avg. SELL} \\\textbf{Precision}\end{tabular} &  \begin{tabular}[h]{c}\textbf{Avg. SELL }\\\textbf{Recall}\end{tabular}
\\
\hline
\begin{tabular}{c}SDCF 1L\end{tabular} & 0.0423 & 0.1778 & 0.0285\\
\hline
\begin{tabular}{c}SDCF 2L\end{tabular} & 0.0650	& 0.1752 & 0.0503\\
\hline
\begin{tabular}{c}SDCF 3L\end{tabular} & 0.0759 & 0.1574 & 0.0635\\
\hline
\begin{tabular}{c}SDCF 4L\end{tabular} & \textbf{0.1410} & \textbf{0.2139} & 0.1250\\
\hline
CNN & 0.0481 & 0.0946 & 0.0379 \\    
%0.0424 & 0.0783 & 0.0348\\
\hline
FCN & 0.0742 & 0.1658 & 0.0802\\
\hline
CNN-TA & 0.0679 & 0.1768 & 0.0487\\
\hline
MFNN & 0.0633 & 0.1034 & \textbf{0.1734}\\
\hline
\end{tabular}
%\end{threeparttable}
\end{table}

%\noindent 
The summary results for individual classes corresponding to every metric are given in Tables \ref{a_summary_buy},~\ref{a_summary_hold}, \ref{a_summary_sell}  above. The average metric values for which the model gives the best performance are average F1 score and precision %and Recall 
for BUY class, average F1 score and recall for HOLD class, average F1 score and precision for SELL class; where F1 score being important metric, as it is the harmonic mean of precision and recall, is the best with our model for all three classes.

\begin{figure}[H]
\begin{tabular}{cc}
\subfloat[CTL 1Layer]{\includegraphics[width = 2.0in, height = 1.8in]{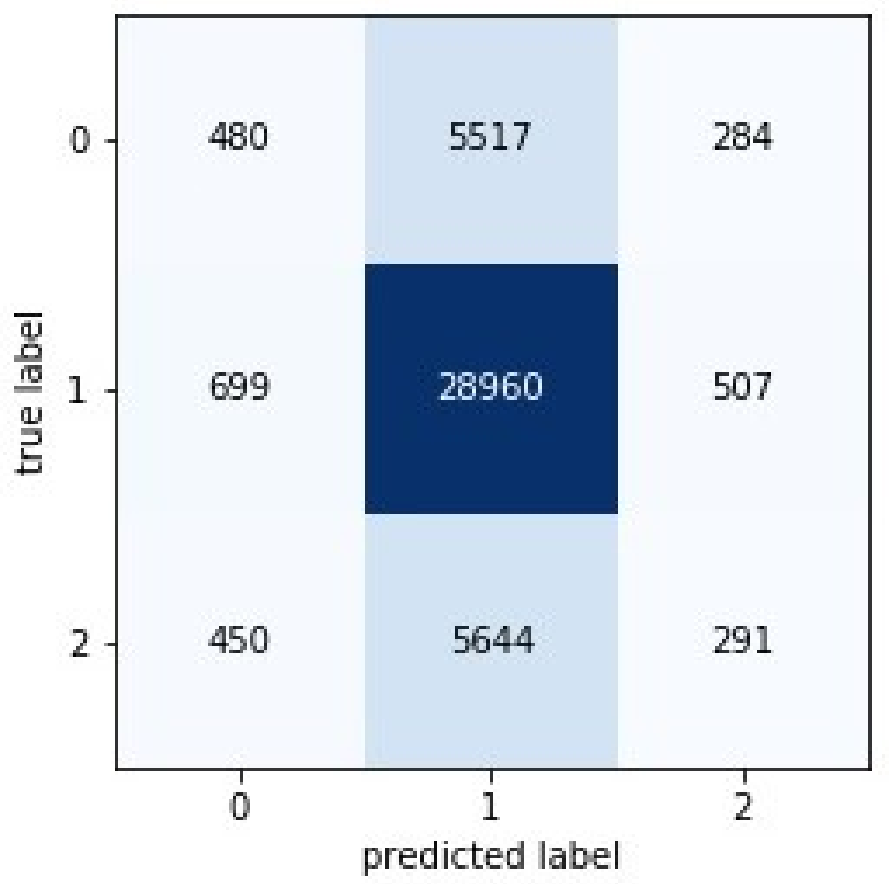}} 
&
\subfloat[CTL 2Layers]{\includegraphics[width = 2.0in, height = 1.8in]{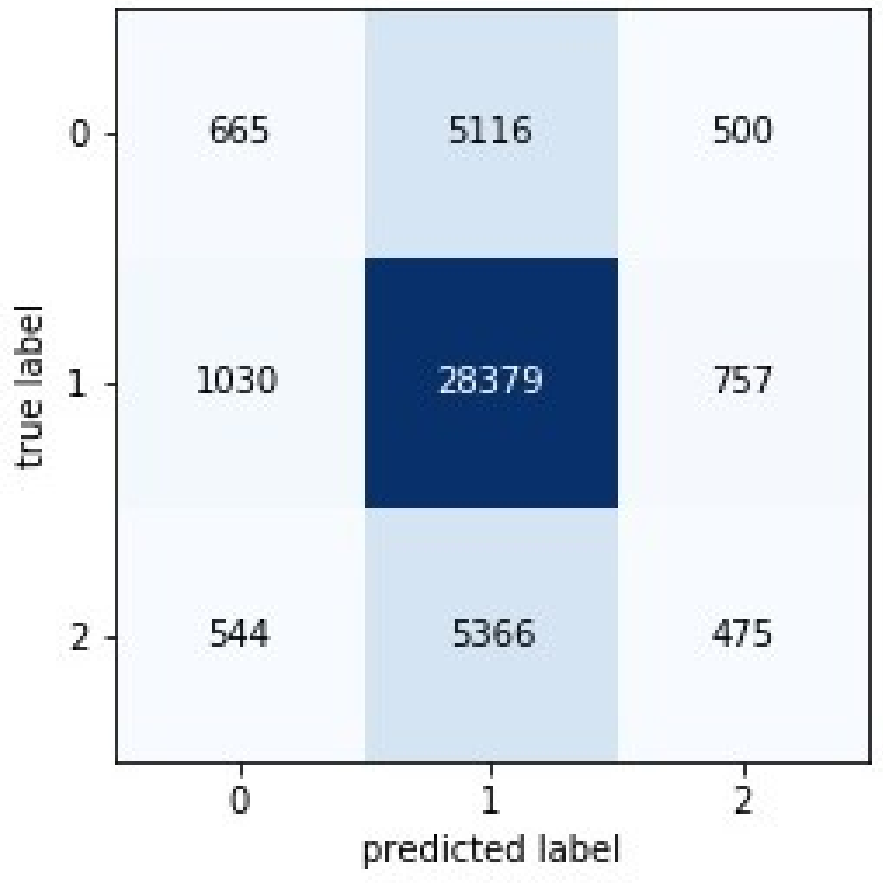}}
\\
\subfloat[CTL 3Layers]{\includegraphics[width = 2.0in, height = 1.8in]{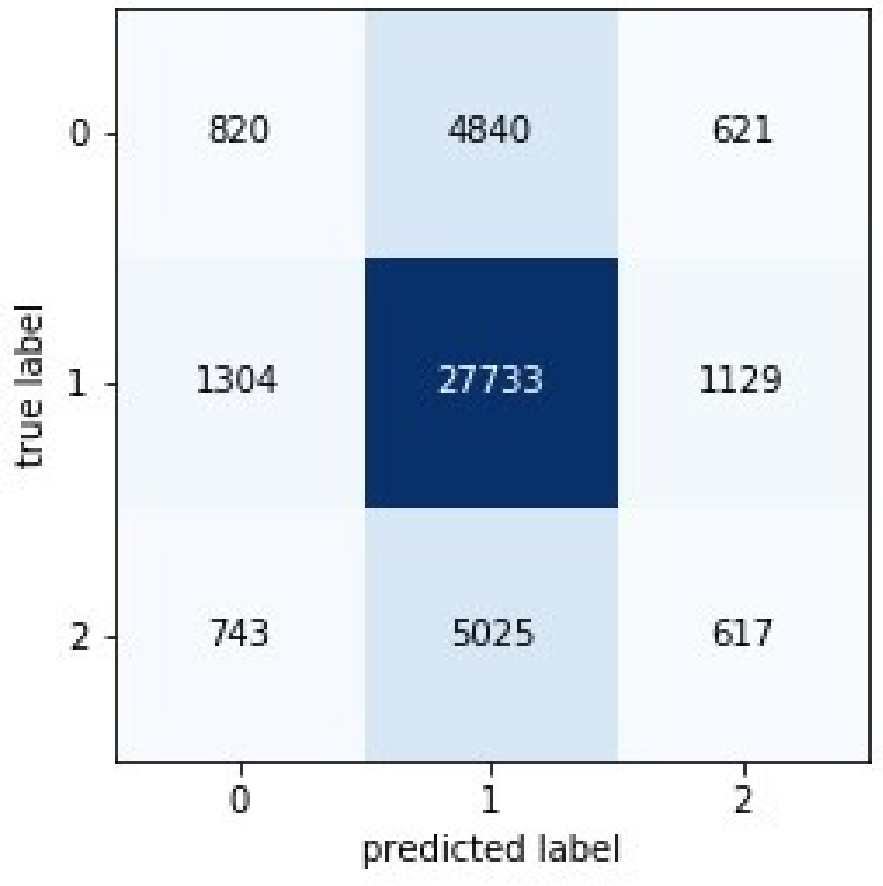}}
&
\subfloat[CTL  4Layers]{\includegraphics[width = 2.0in, height = 1.8in]{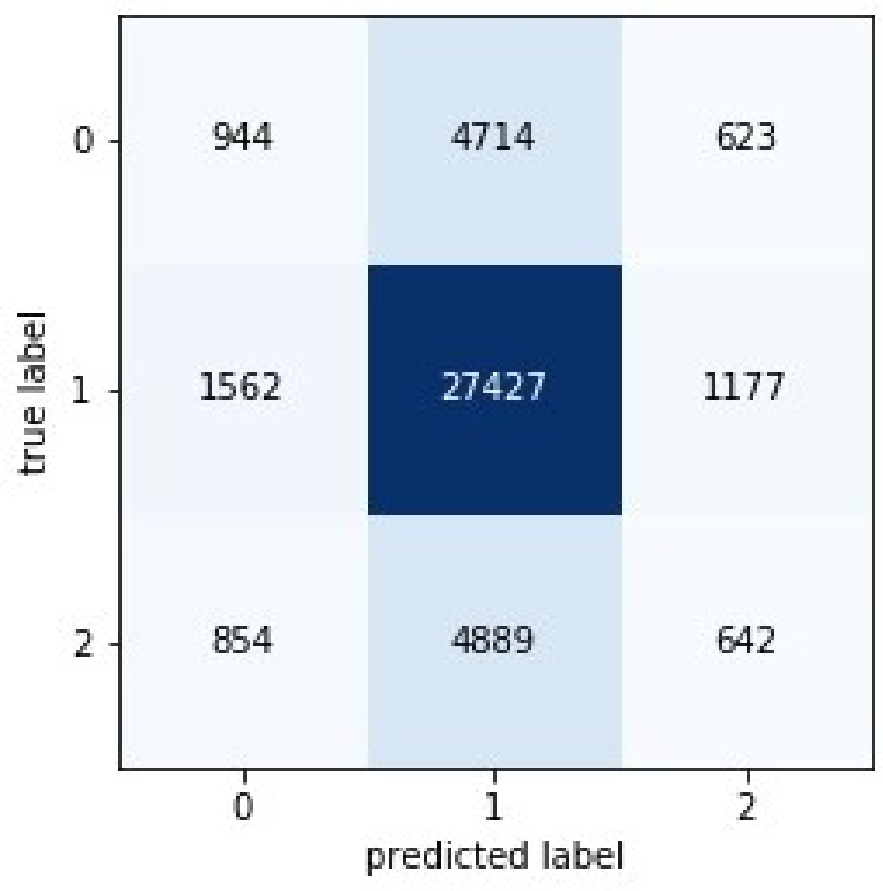}}
\end{tabular}
\caption{Confusion matrices corresponding to the different number of CTL layers of the architecture: a) 1 layer of CTL (shallow version), b) 2 layers of CTL (deep version), c) 3 layers of CTL (deep version) and d) 4 layers of CTL (deep version) where 0 - BUY, 1 - HOLD, 2 - SELL signals.} \label{confusion_matrices}
\end{figure}

%and similar performance for average F1 scores for BUY and SELL classes and average Precision for SELL class under the same err\_dif of 0.02. 
%really well versus the other two BUY and SELL classes from F1 score metric. 
As we can observe, the performance for HOLD class decrease when increasing the number of layers for our model. However, we can also see that there is an increase in correct identification for BUY and SELL points despite the fact that BUY and SELL points appear extremely less in case of every stock as compared to HOLD points. The latter identification capacity is actually more crucial for the financial system as it directly influences the financial gains or loss. Moreover, the overall individual class performance indicate that the model captures all three classes i.e. BUY, HOLD and SELL well. %therefore, capturing the BUY and SELL points correctly is difficult and important.}
%which the model correctly identifies with an increase in CTL layers. }
%Although, the performance is not the best as compared with the benchmarks but still gives good performance as also suggested from the results. 
This is also indicated in the confusion matrices, given for each of the shallow and deeper versions of our framework in Figure \ref{confusion_matrices}. With an increase in layers, the model starts to more correctly identify the BUY and SELL points. The HOLD signal has more false positives with shallow architecture (SDCF 1L) that decreases with the increase in layer number, which is important for the system in order to correctly classify other class points. Additionally, the overall performance with our model is better than the CNN.

%\noindent 
To better analyze the framework performance, we calculate the weighted F1 score, precision and recall metric values for all the stocks under consideration. We calculate the weighted values to incorporate the class imbalance for every stock. The detailed and summary results are given in Table \ref{a_weighted_detailed} in \ref{appendix_a} and Table \ref{a_weighted_summary}. Again, the results comprise two sets of values marked in bold or red with the same err\_dif of 0.02. There are 6, 9, and 5 stocks with respect to the metrics F1 score, precision and recall for which the model performs greater than or equal to the performance given by the state-of-the-arts. Also, there are 6, 3 and 6 stocks for the metrics F1 score, precision and recall respectively for which the model gives the next best performance under 0.02 err\_dif. Although the BUY and SELL classes performance with the 4 CTL Layers deep architecture is better than the benchmarks compared against, but the overall performance from the average weighted metric is suggestive of the good performance with the 3 layers deep architecture classification wisely. This is also suggested from the financial results explained later. 

Again analyzing explicitly for CNN, we have 4, 2 and 7 stocks with greater than or equal performance; and 3, 2 and 3 stocks under similar/next best performance for the metrics F1 score, precision and recall respectively. As can be referenced from the statistics presented here, our model is giving better results with greater than or equal and the next best/similar performances except for the number of stocks for recall metric are slightly more with CNN under greater than or equal to performance. However, the next best performance statistic for the recall metric is much better than CNN. Overall performance on an average is good with our proposed model as compared to the benchmarks and CNN which can be also referred from Table \ref{a_weighted_summary}. For a deeper understanding of the aforementioned statistics, please refer to Table \ref{comparison} in ~\ref{appendix_c}. 

\begin{table}[H]
%\begin{threeparttable}
\caption{\textbf{Summary of Weighted Classification Results for Stock Trading}}\label{a_weighted_summary}
\centering
\begin{tabular}{|c|c|c|c|}
\hline
\textbf{Method} &  \begin{tabular}[h]{c}\textbf{Avg.} \\ \textbf{F1 Score}\end{tabular} &  \begin{tabular}[h]{c}\textbf{Avg. }\\ \textbf{Precision}\end{tabular} &  \begin{tabular}[h]{c}\textbf{Avg.} \\ \textbf{Recall}\end{tabular} 
\\
\hline

\begin{tabular}{c}SDCF 1L\end{tabular} & 0.6169 & \textbf{0.6216} & \textbf{0.6941}\\
\hline
\begin{tabular}{c}SDCF 2L\end{tabular} & 0.6229 & 0.6207 & 0.6867\\
\hline
\begin{tabular}{c}SDCF 3L\end{tabular} & \textbf{0.6250} & 0.6146 & 0.6784\\
\hline
\begin{tabular}{c}SDCF 4L\end{tabular} & 0.5345 & 0.5464 & 0.5890\\
\hline
CNN & 0.6182 & 0.5907 & 0.6898 \\
\hline
FCN & 0.6090 & 0.6079 & 0.6725\\
\hline
CNN-TA & 0.6148 & 0.6161 & 0.6575\\
\hline
MFNN & 0.4162 & 0.5509 & 0.4676\\
\hline
\end{tabular}
%\end{threeparttable}
\end{table}
\subsection{Financial Analysis}

%\noindent 
It is very important to analyse the performance from a financial perspective to understand the quality of predictions made by our model. For this, as explained earlier, we have calculated the AR values with the predictions generated by each of the techniques for every stock over 11 years. We also calculate the AR values with the True labels for every stock over the same period. Finally, we calculate the absolute difference/error between the AR values from Predictions and the AR values from True labels. We average the absolute difference values for all stocks yielding the so-called Mean Absolute Error. The detailed results are given in Table \ref{b_detailed}. With our proposed model 5 stocks have the best performance whereas with CNN-TA there is 1 stock and 2 stocks under MFNN and FCN. On the whole, the performance is good with our proposed model as also evident from the summary results in Table \ref{b_summary} where we have a mean of the absolute difference/error(MAE) between the True AR and Predicted AR. Also, there are 3 stocks for which the proposed model gives an equal performance as the other benchmark techniques. Here, this set of results is illustrating that, despite the higher capability of identifying the BUY and SELL points with 4 layers deep CTL, the AR values are better predicted with the 3 layers deep CTL framework.  

With respect to CNN, there are only 2 stocks for which CNN performs better than any benchmarks and our proposed models, and 3 stocks for which it gives an equal performance. Thus, from the combined (greater than or equal to and next best / similar), average and the financial results, the CNN results are less performant than our model. This also indicates that the quality of predictions made with our model is better than CNN as the identified class labels give AR values quite close to the True AR values. This remains true for all the benchmarks. The statistics presented here can be deduced from Table \ref{comparison} in ~\ref{appendix_c} for complete understanding.
%Also, if we analyze overall Classification weighted metric results, then our model performs better than the CNN.
%and 1 stock for which it gives next best performance under 0.02 err\_diff. 
%that despite the number of stocks qualifying for the good performance with our model are a bit less than the numbers with CNN computationally, 

\begin{table}[H]
%\begin{threeparttable}
\caption{\textbf{Summary of Financial Results for Stock Trading}}\label{b_summary}
\centering
\begin{tabular}{|c|c|}
\hline
\textbf{Method} &  \begin{tabular}[h]{c} \textbf{MAE AR}\end{tabular} 
\\
\hline
\begin{tabular}{c}SDCF 1L\end{tabular} & 22.5613
\\
\hline
\begin{tabular}{c}SDCF 2L\end{tabular} & 20.7227\\
\hline
\begin{tabular}{c}SDCF 3L\end{tabular} & \textbf{20.5067}\\
\hline
\begin{tabular}{c}SDCF 4L\end{tabular} & 22.8287\\
\hline
CNN & 21.1140\\
\hline
FCN & 23.7720\\
\hline
CNN-TA & 22.1380\\
\hline
MFNN & 22.3040\\
\hline
\end{tabular}
%\end{threeparttable}
\end{table}

%\noindent 

To further understand the better supervised learning for both regular CNN and our SDCF framework, we visualize channel-wise $X_c$ features for both the frameworks which are obtained after the last maxpool layer for the 3 convolutional layers deep framework. The following Figure \ref{visuals_x} shows the visualizations of the features for one sample of the stock `BSELINFRA.BO'.

\begin{figure}[H]
\begin{tabular}{ccccc}
%\subfloat[CTL 1Layer]
\multicolumn{5}{c}{\footnotesize Features generated by the proposed \emph{SDCF} network.}
\\
\includegraphics[width = 1.0in, height = 2.5in]{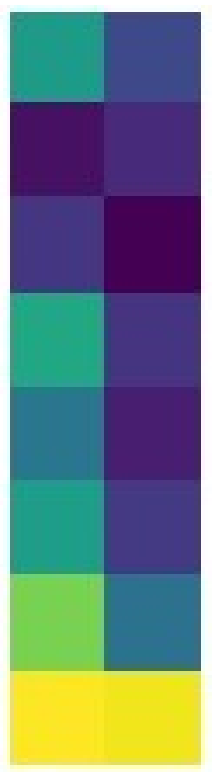}
&
%\subfloat[CTL 2Layers]
\includegraphics[width = 1.0in, height = 2.5in]{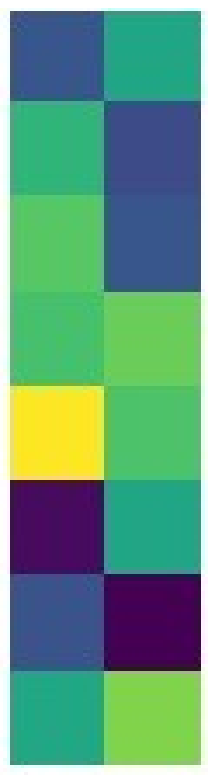}
&
%\subfloat[CTL 2Layers]
\includegraphics[width = 1.0in, height = 2.5in]{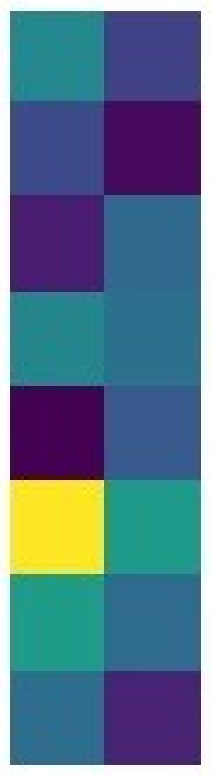}
&
%\subfloat[CTL 2Layers]
\includegraphics[width = 1.0in, height = 2.5in]{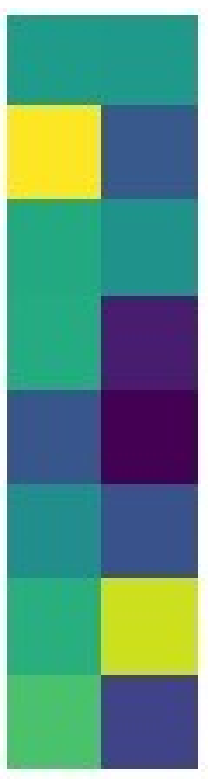}
&
%\subfloat[CTL 2Layers]
\includegraphics[width = 1.0in, height = 2.5in]{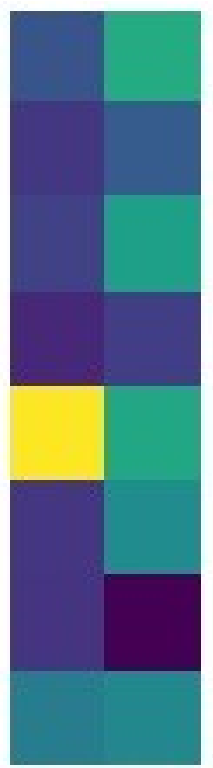}
\\[1em]

\multicolumn{5}{c}{\footnotesize Features generated by a standard CNN with a similar architecture.}
\\[-1em]
\subfloat[Channel $X_1$  Close Price]{\includegraphics[width = 1.0in, height = 2.5in]{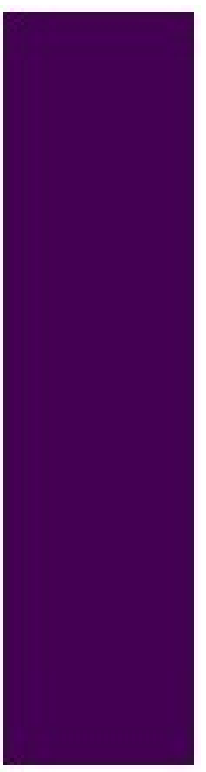}}
% \subfloat[Channel $X_1$ \newline \centering{Close Price}]{\includegraphics[width = 1.0in, height = 2.54in]{CNN_X1.eps}}
&
\subfloat[Channel $X_2$ Open Price]{\includegraphics[width = 1.0in, height = 2.5in]{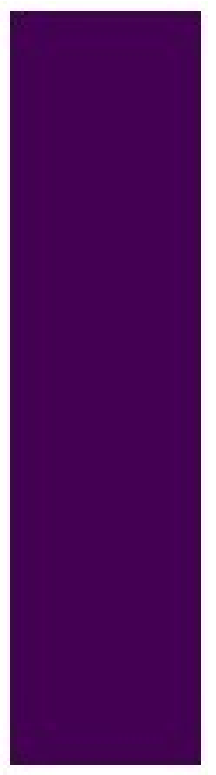}}
&
\subfloat[Channel $X_3$ High Price]{\includegraphics[width = 1.0in, height = 2.5in]{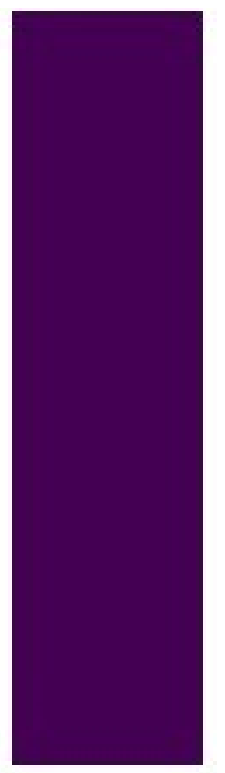}}
&
\subfloat[Channel $X_4$ Low Price]{\includegraphics[width = 1.0in, height = 2.5in]{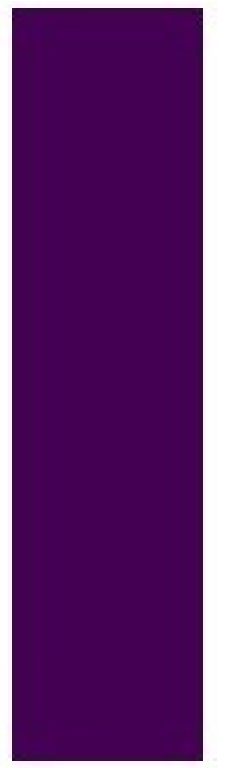}}
&
\subfloat[Channel $X_5$ NAV]{\includegraphics[width = 1.0in, height = 2.5in]{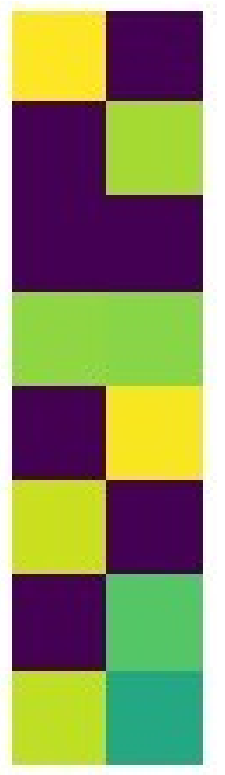}}%_newest.eps}}

\end{tabular}
\caption{Visualization of channel-wise features $X_c$ for SDCF  versus a standard CNN, for one sample of stock BSELINFRA.BO (with 16x1 as the shape of the features obtained and resized to 8x2 for better visualization)}\label{visuals_x}
\end{figure}
\newpage
As can be seen from Figure \ref{visuals_x}, heatmap for each channel corresponding to the prices(Close, Open, High and Low) show no variation in the case of CNN as compared to the SDCF architecture. While it shows some variations for the features learnt corresponding to NAV, however, the features are still better learnt with SDCF. Also, darker the color in the heatmap, more it is indicative of the larger negative exponent values. In the case of CNN, hence, the values are very very small that are almost diminishing to zero. This also corroborates the fact that the filters learnt with our model are distinct due to the ``log-det" term added which further gives different features with very less redundancy. Thus, the visualizations of these channel-wise features are also supportive of better supervised training with our framework than CNN.  

\begin{figure}[H]
\begin{tabular}{cc}
%\subfloat[CTL 1Layer]
\includegraphics[width = 2.1in, height = 1.7in]{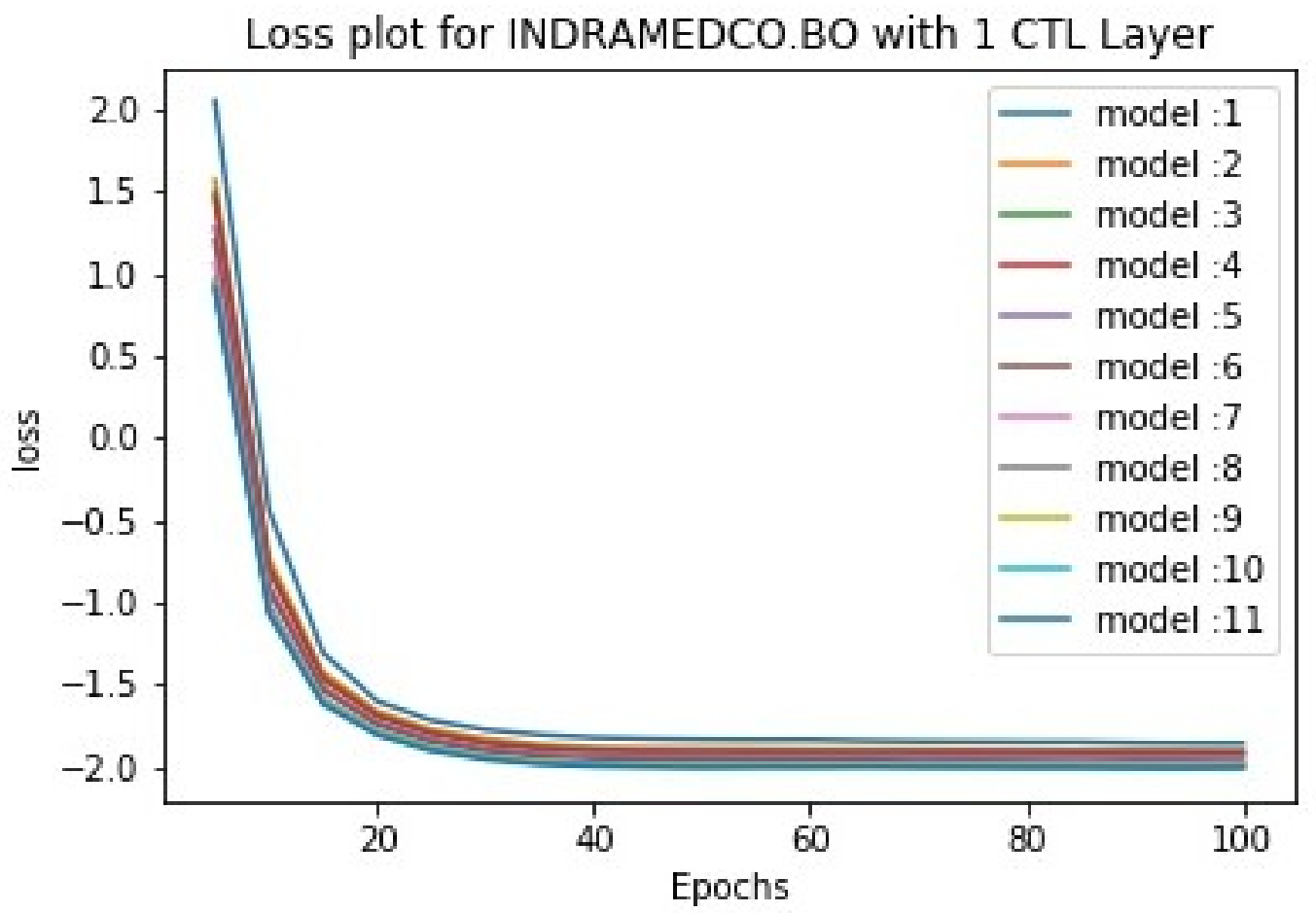}
&
%\subfloat[CTL 2Layers]
\includegraphics[width = 2.1in, height = 1.7in]{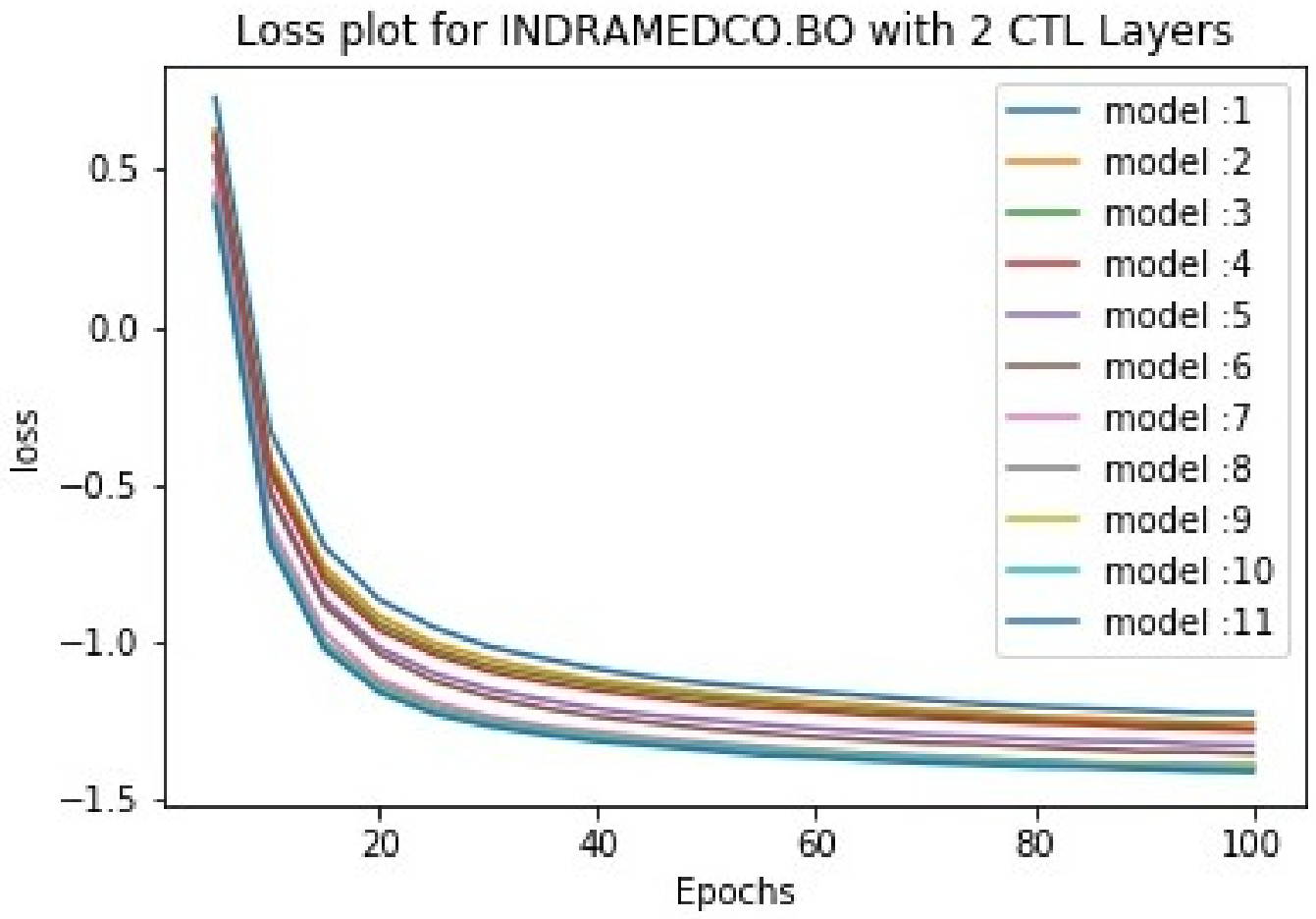}
\\
\subfloat[CTL 1 layer]{\includegraphics[width = 2.1in, height = 1.7in]{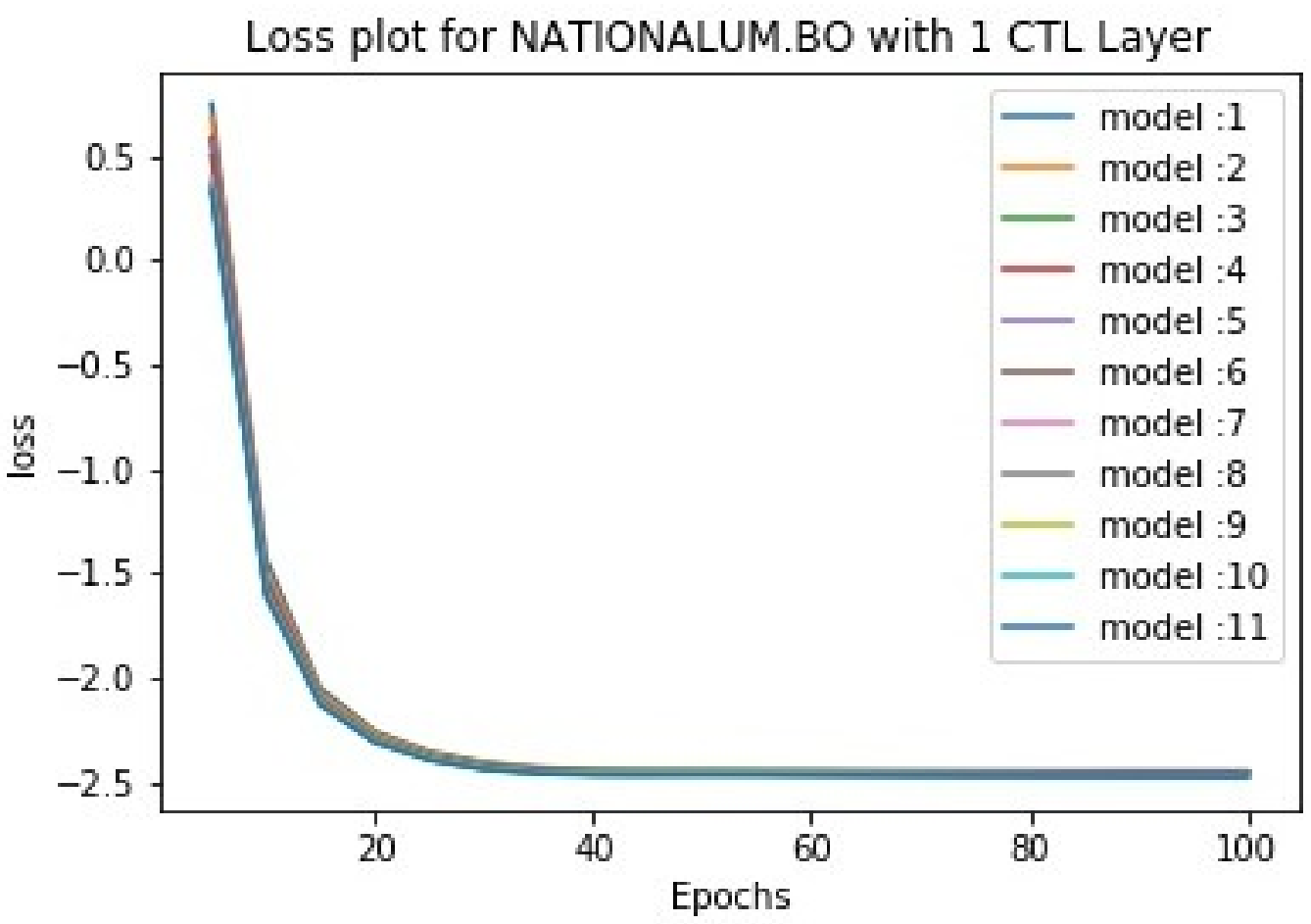}} 
&
\subfloat[CTL 2 layers]{\includegraphics[width = 2.1in, height = 1.7in]{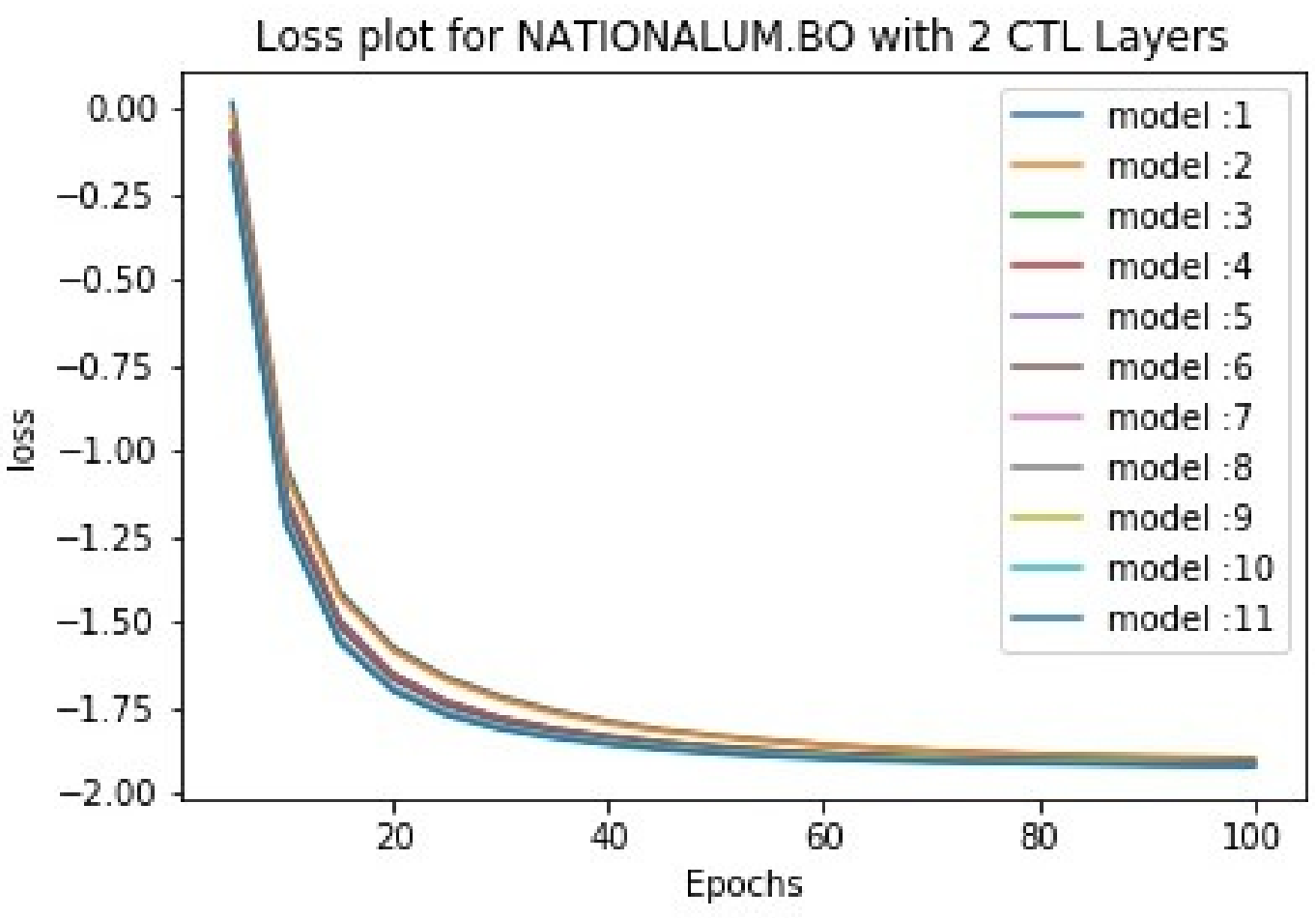}}
\\
%\subfloat[CTL 3Layers]
\includegraphics[width = 2.1in, height = 1.7in]{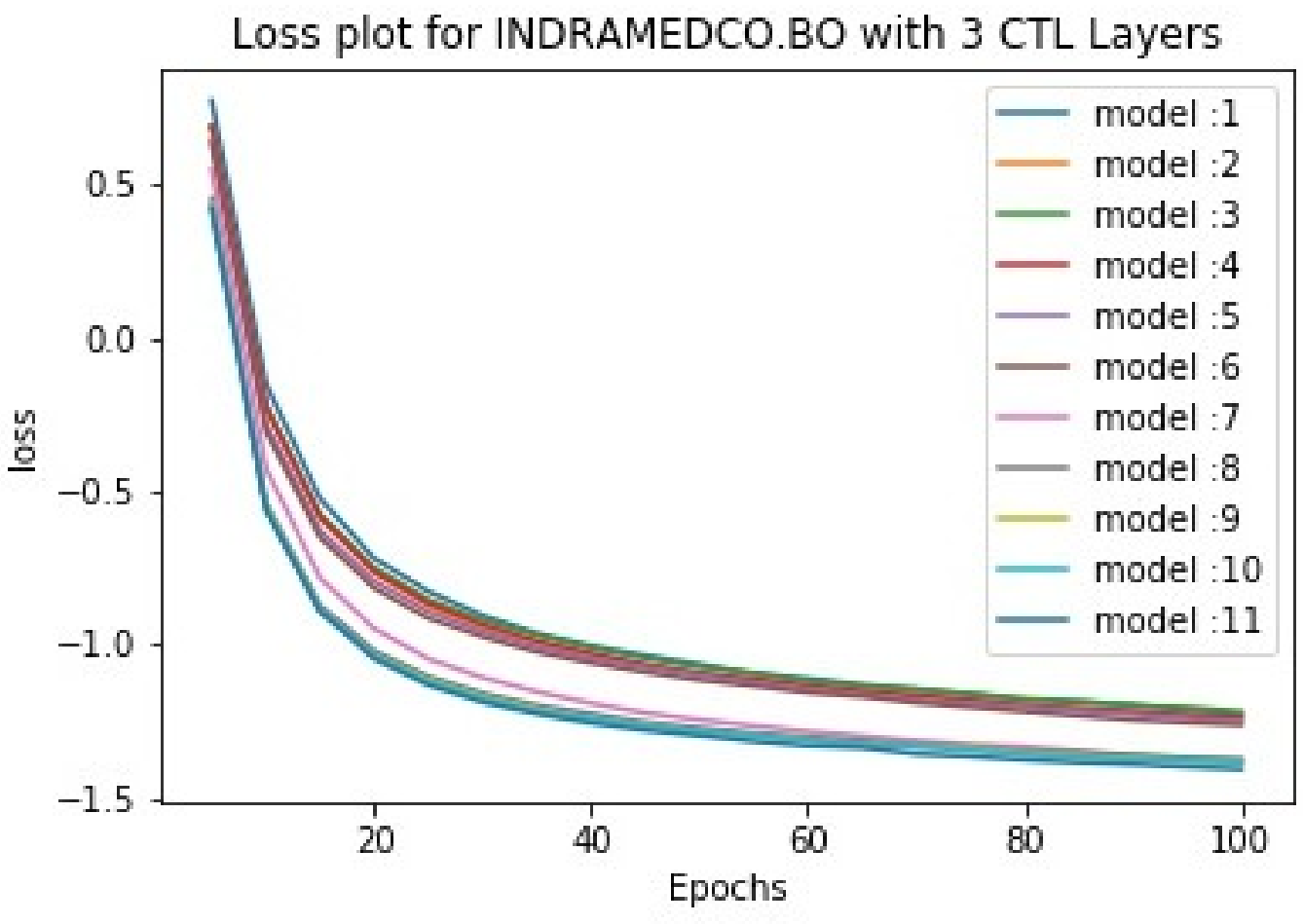}
&
\includegraphics[width = 2.1in, height = 1.7in]{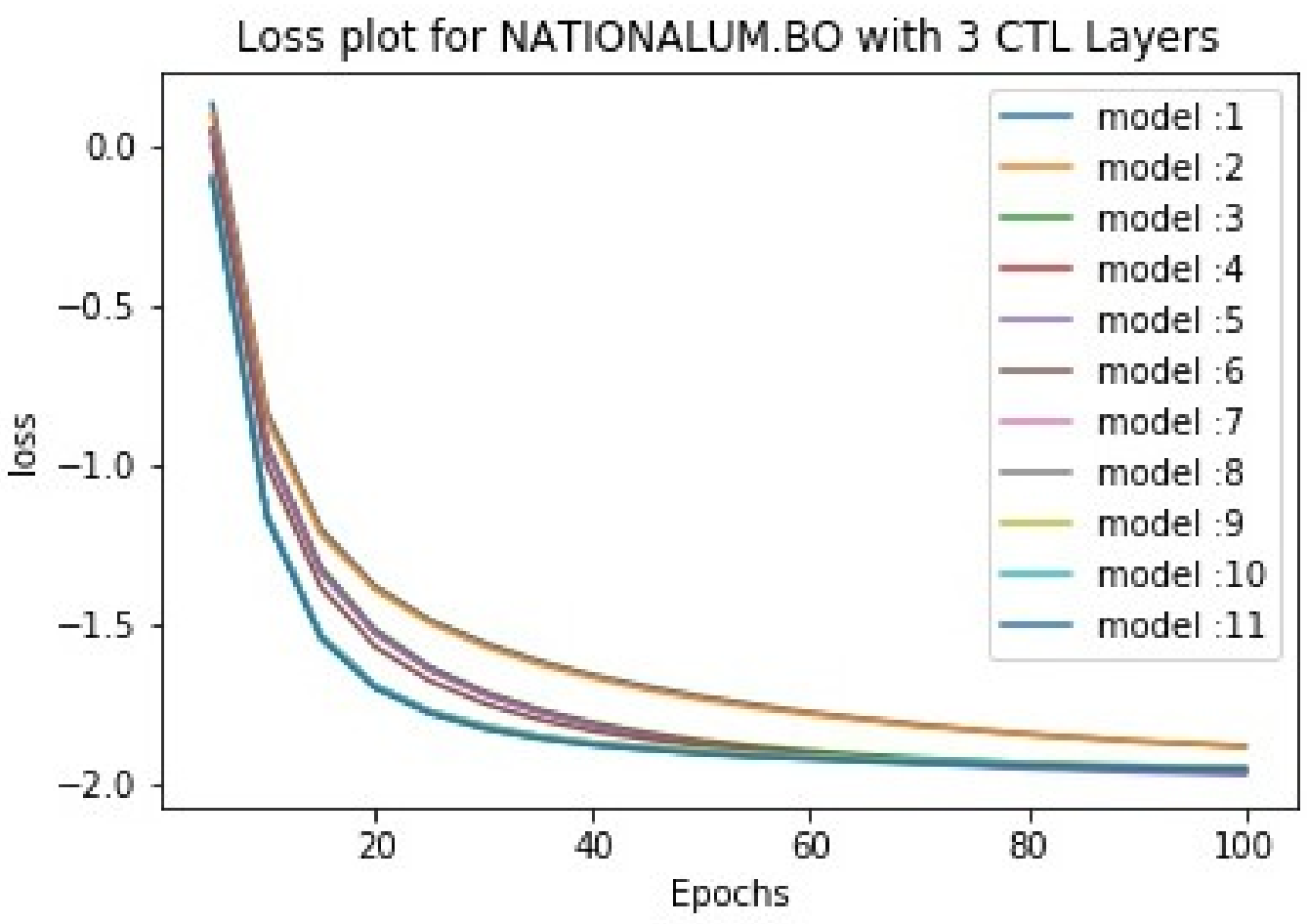}
\\
\subfloat[CTL 3 layers]{\includegraphics[width = 2.1in, height = 1.7in]{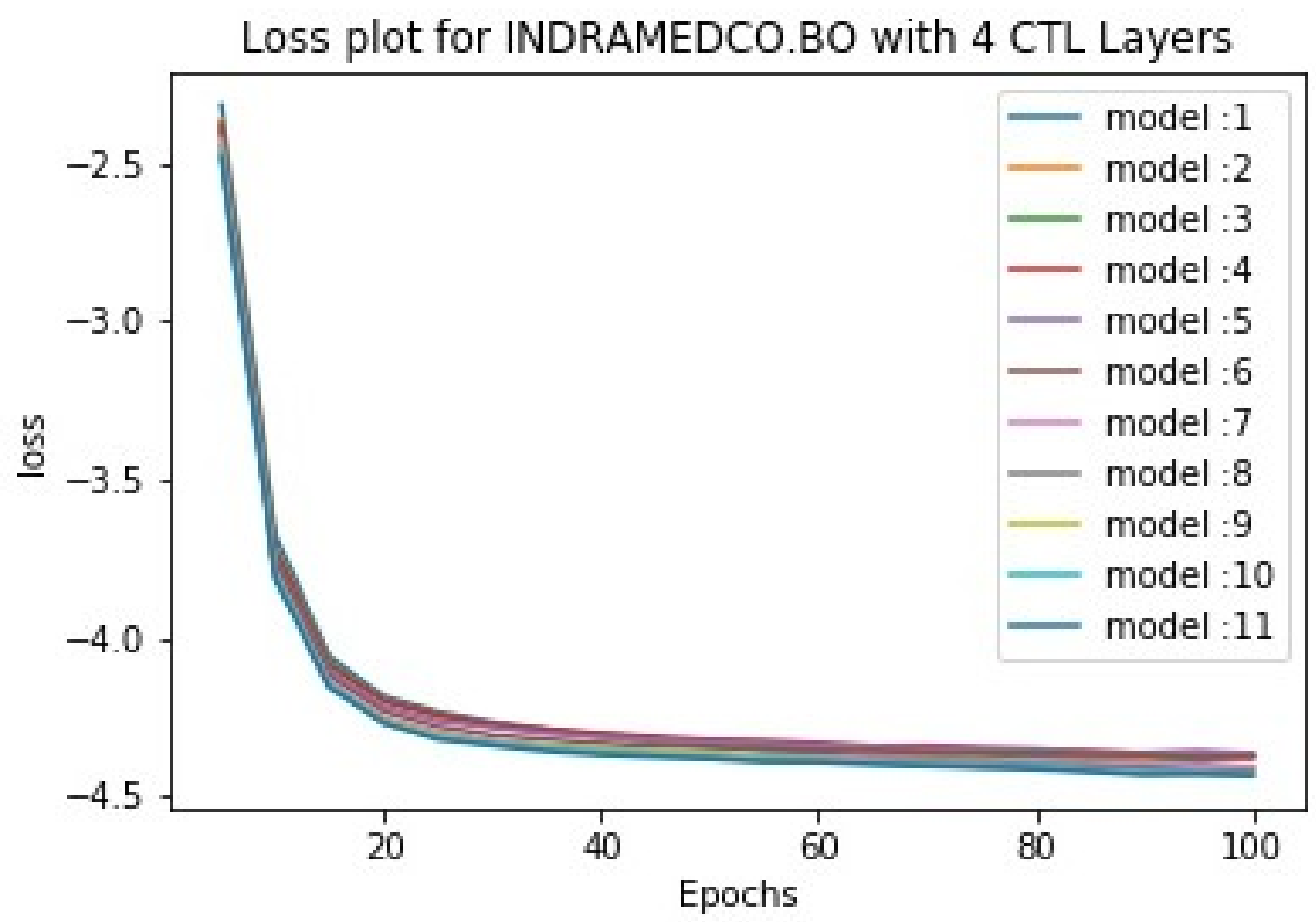}}
&
\subfloat[CTL 4 layers]{\includegraphics[width = 2.1in, height = 1.7in]{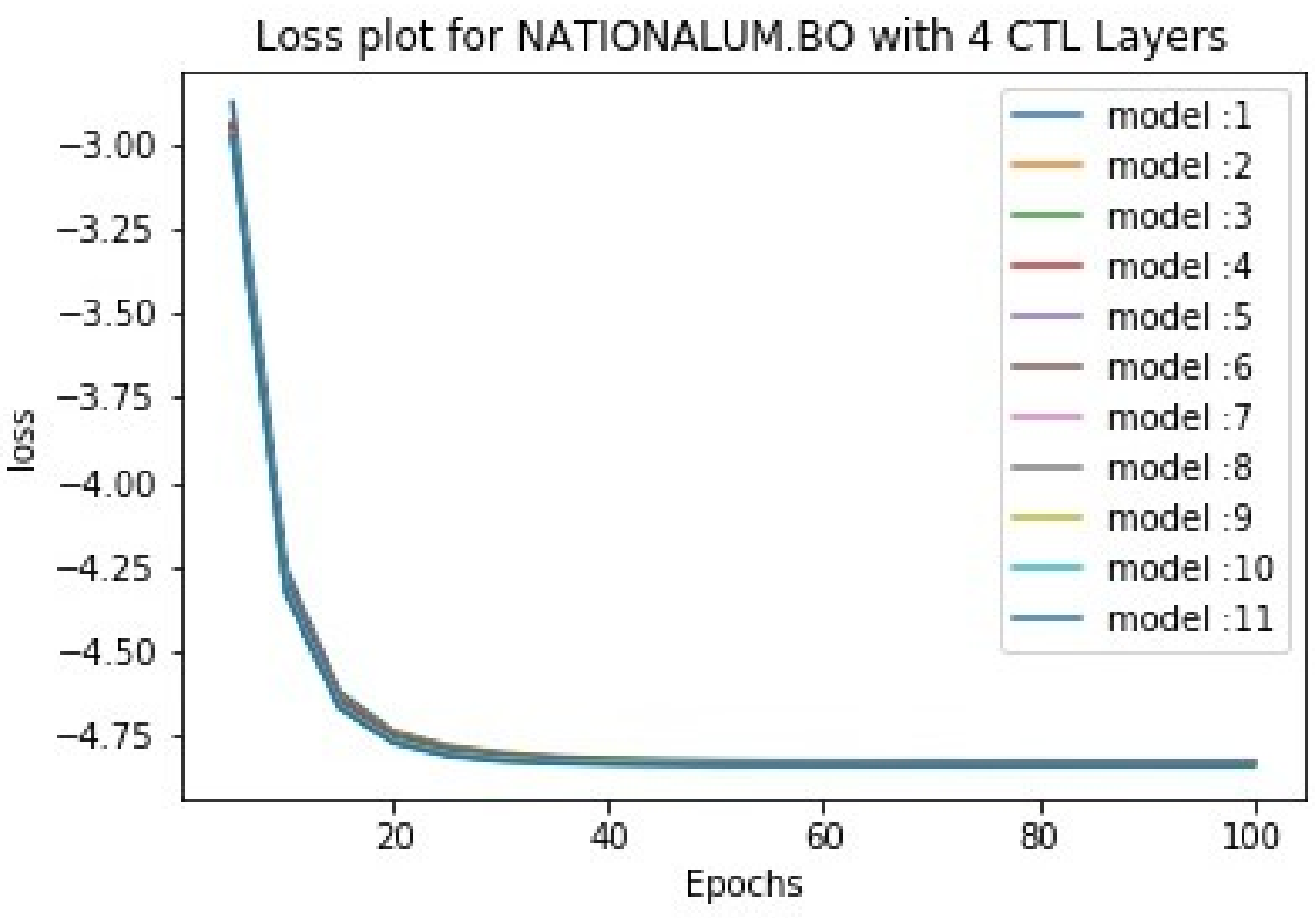}}
\end{tabular}
\caption{Evolution of the loss during training for few stock examples, of our proposed model with (a) CTL 1 layer, (b) CTL 2 layers, (c) CTL 3 layers and (d) CTL 4 layers. 
}\label{loss_plots}
\end{figure}

%Putting it all together, 
In order to test our architecture's capability further, we have performed experiments for two additional window sizes, namely 5 and 20. In order to avoid extensive space utilization, we present here only the comparative summary results - Weighted F1 Score(Classification Metric) and MAE AR(Financial metric) in Table \ref{weighted_summary_5_10_20} for window sizes  5 and 20 along with the summarized results for window size 10. Our method yields the best results on an aggregate. Even though CNN-TA yields better AR for a solo case (window size 20), it does not reach better results in terms of weighted F1 for the same scenario. Furthermore, CNN-TA cannot be run for all small window sizes (such as 5), hence cannot be deemed as an all-purpose go-to method. Small window sizes are crucial for highly non-stationary stocks and the inability of a method to handle such stocks is a major shortcoming.
%  However, due to the inherent architecture of one of the benchmark techniques - CNN-TA, we could not execute it for the window size of 5 for the dataset. The detailed results for window size 10 are present in Appendices A, B and C.
Overall, our model performs better than benchmarks and CNN both classification-wise and financially, specifically, it gives the best performance with 3 CTL layers deep SDCF framework of all the 4 SDCF architectures. We also display the empirical convergence plots for a few stocks, namely INDRAMEDCO.BO and NATIONALUM.BO in Figure \ref{loss_plots} for both shallow and deeper versions. We can see that the training loss decreases to a point of stability for each example  considered.

\begin{table}[H]
%\begin{threeparttable}
\caption{\textbf{Comparative Summary Results for Stock Trading for window sizes 5,10,20}}\label{weighted_summary_5_10_20}
\centering
\begin{tabular}{|c|c|c|c|c|c|c|}
\hline
\multirow{2}{*}{\textbf{Method}} &  \multicolumn{2}{c|}{\textbf{Window Size 5}} & \multicolumn{2}{c|}{\textbf{Window Size 10}} &
\multicolumn{2}{c|}{\textbf{Window Size 10}} \\
\cline{2-7}
& \textbf{F1} & \textbf{MAE AR} & \textbf{F1} & \textbf{MAE AR}
& \textbf{F1} & \textbf{MAE AR} \\
\hline
\begin{tabular}{c}SDCF 1L\end{tabular} & 0.6141 & 22.4947 &  0.6169 & 22.5613 & 0.6194 & 22.4453
\\
\hline
\begin{tabular}{c}SDCF 2L\end{tabular} & 0.6148 & 24.3820 &  0.6229 & 20.7227 & 0.6242 & 25.0200\\
\hline
\begin{tabular}{c}SDCF 3L\end{tabular} & \textbf{0.6207} &  \textbf{20.9193} & \textbf{0.6250} & \textbf{20.5067} & \textbf{0.6262} & 25.7667\\
\hline
\begin{tabular}{c}SDCF 4L\end{tabular} & 0.6157 & 21.5427 & 0.5345 & 22.8287 & 0.6254 & 26.1007\\
\hline
CNN & 0.6095 & 22.0113 & 0.6182 & 21.1140 & 0.6217 & 22.9560\\
\hline
FCN & 0.6131 & 23.3107 & 0.6090 & 23.7720 & 0.6120 & 24.2233\\
\hline
CNN-TA* & - & - & 0.6148 & 22.1380 & 0.6246 & \textbf{20.3820} \\
\hline
MFNN & 0.4105 & 23.4820 & 0.4162 & 22.3040 & 0.4869 & 23.2620\\
\hline
\end{tabular}
\begin{tablenotes}
\item \small{*CNN-TA cannot be run for window size 5 due to its inherent structure}
\end{tablenotes}
%\end{threeparttable}
\end{table}

\section{Conclusion}
%\noindent 
In this work, we propose SDCF, a deep fusion end-to-end framework for the processing of stock trading data. Unlike other deep learning models, our framework is a fusion supervised framework. It relies on a novel deep version of our recently proposed CTL model. We have applied the proposed model for stock trading leading to very good performance. In particular, the classification results are better with the proposed SDCF model, than with the 1-D CNN approach. Also, the features $X_c$ visualized for each channel and each method indicate the better feature learning with SDCF. The results show that the proposed solution is superior to CNN and other state-of-the arts techniques in this problem.

We believe that the framework is generic enough to handle other multi-channel fusion problems as well. In the future, we plan to extend the application to other fusion 1-D as well as 2-D multi-channel problems to test its generality. For example, we plan to implement it in the biomedical domain to analyse PPG and ECG signals for the blood pressure estimation pertaining to the 1-D multi-channel problem. In case of 2-D problems, we would like to do multi-spectral image classification using this technique. Currently, the shortcoming with our model is that it takes slightly more time than CNN, for its training. Thus, we will investigate on the reduction of the time complexity of our framework in order to make it more efficient from this viewpoint.

The current purpose of our paper is to introduce our new algorithm and to show by means of several experiments that it is an effective tool for predicting stocks. However, stock price prediction may be seen as a too rudimentary problem in financial analytics. As a next step, we would like to investigate the use of our algorithm to study if it can emulate (human) expert-like suggestions. For example, fund managers suggest `buy stock XYZ at a price ABC' or `sell stock ZYX at price CBA'. We would like to see if our algorithm can make such predictions given a time horizon. If possible, we would like to extend the algorithm to emulate more abstract financial operations such as `hedging (longs and shorts)'.
%However, the only limitation with the current work is that the maxpooling is not supported if we extend our model beyond 3 layers and hence only SELU activation can be applied. However, this is dependent on the input sequence length considered which in our case is less due to which we are restricted to employ maxpooling till $3$ layers  

\section{Acknowledgement}
\noindent This work was supported by the CNRS-CEFIPRA project under grant NextGenBP PRC2017.

% \section{Bibliography styles}

% There are various bibliography styles available. You can select the style of your choice in the preamble of this document. These styles are Elsevier styles based on standard styles like Harvard and Vancouver. Please use Bib\TeX\ to generate your bibliography and include DOIs whenever available.

% Here are two sample references: \citep{Feynman1963118,Dirac1953888}.

% \section*{References}
\newpage
\bibliography{mybibfile}

\newpage
\appendix
\section{Class-Wise Classification Results for Stock Trading}
\label{appendix_a}
\noindent This section displays all the tables with the Classification Metrics results, both class-wise and weighted, for stock trading. 

\begin{longtable}[!htb]{|c|c|c|c|c|}
\caption{\textbf{Classification Results for BUY Class for Stock Trading}}\label{a_detailed_buy}\\
\hline
\textbf{SYMBOL} & \textbf{Method} & 
\begin{tabular}[h]{c}
\textbf{BUY} \\
\textbf{F1 Score}
\end{tabular}
& 
\begin{tabular}[h]{c}
\textbf{BUY} \\
\textbf{Precision}
\end{tabular}
& 
\begin{tabular}[h]{c}
\textbf{BUY} \\
\textbf{Recall}
\end{tabular}
\\
\hline
\textbf{ALKYLAMINE.BO} &
\begin{tabular}{c}SDCF 1L\end{tabular}&0.1022 & 0.2450 & 0.0646\\
&\begin{tabular}{c}SDCF 2L\end{tabular}&0.1708 & 0.2222 & 0.1387\\
&\begin{tabular}{c}SDCF 3L\end{tabular}&0.1670&0.2056&0.1407\\
& \begin{tabular}{c}
SDCF 
 4L 
\end{tabular}  & \textbf{0.1880} & \textbf{0.2500} & \textbf{0.1507}\\
%& \cred{0.1880} & 0.2500 & \cred{0.1507}\\
&CNN & 0.1678 & 0.2458 & 0.1274\\
%\textbf{0.1991} & \textbf{0.2890} & \textbf{0.1518} \\
&FCN&0.1131&0.2484&0.0732\\
\hline
\end{longtable}

\clearpage

\begin{longtable}[!htb]{|c|c|c|c|c|}
\caption{\textbf{Classification Results for HOLD Class for Stock Trading}}\label{a_detailed_hold}\\
\hline
\textbf{SYMBOL} & \textbf{Method} & 
\begin{tabular}[h]{c}
\textbf{HOLD}\\
\textbf{F1 Score}
\end{tabular}
& 
\begin{tabular}[h]{c}
\textbf{HOLD} \\
\textbf{Precision}
\end{tabular}
& 
\begin{tabular}[h]{c}
\textbf{HOLD} \\
\textbf{Recall}
\end{tabular}
\\
\hline
\textbf{ALKYLAMINE.BO}&\begin{tabular}{c}SDCF 1L\end{tabular}&0.7216&0.6062&0.8912\\
&\begin{tabular}{c}SDCF 2L\end{tabular}&0.6932&\cred{0.6137}&0.7963\\
&\begin{tabular}{c}SDCF 3L\end{tabular}&0.6762&0.6113&0.7565\\
& \begin{tabular}{c}
SDCF 
 4L 
\end{tabular}  & 0.5761 & 0.5072 & 0.6667\\
& CNN & 0.7214 & \textbf{0.6196} & 0.8632 \\
&FCN&\textbf{0.7318}&0.6060&0.9236\\
\hline
\end{longtable}

\begin{longtable}[!htb]{|c|c|c|c|c|}
\caption{\textbf{Classification Results for SELL Class for Stock Trading}}\label{a_detailed_sell}\\
\hline
\textbf{SYMBOL} & \textbf{Method} & 
\begin{tabular}[h]{c}
\textbf{SELL}\\
\textbf{F1 Score}
\end{tabular}
& 
\begin{tabular}[h]{c}
\textbf{SELL} \\
\textbf{Precision}
\end{tabular}
& 
\begin{tabular}[h]{c}
\textbf{SELL} \\
\textbf{Recall}
\end{tabular}
\\
\hline
\textbf{ALKYLAMINE.BO}&\begin{tabular}{c}SDCF 1L\end{tabular}&0.1207&0.2279&0.0821\\
&\begin{tabular}{c}SDCF 2L\end{tabular}&0.1820&0.2681&0.1378\\
&\begin{tabular}{c}SDCF 3L\end{tabular}&0.1945&0.2522&0.1583\\
& \begin{tabular}{c}
SDCF 
 4L 
\end{tabular}  & \textbf{0.3594} & \textbf{0.3710} & 0.3485\\
& CNN & 0.1278 & 0.2537 & 0.0854 \\
&FCN&0.0879&0.2800&0.0521\\
\hline
\end{longtable}

\clearpage
\begin{longtable}[!htb]{|c|c|c|c|c|}
\caption{\textbf{Weighted Classification Results for Stock Trading }}\label{a_weighted_detailed}\\
\hline
\textbf{SYMBOL} & \textbf{Method} & 
\begin{tabular}[h]{c}
\textbf{F1 Score}
\end{tabular}
& 
\begin{tabular}[h]{c}
\textbf{Precision}
\end{tabular}
& 
\begin{tabular}[h]{c}
\textbf{Recall}
\end{tabular}
\\
\hline
\textbf{ALKYLAMINE.BO}&\begin{tabular}{c}SDCF 1L\end{tabular}&0.4738&0.4559&0.5590\\
&\begin{tabular}{c}SDCF 2L\end{tabular}&\cred{0.4824}&\cred{0.4635}&0.5278\\
&\begin{tabular}{c}SDCF 3L\end{tabular}&0.4741&0.4554&0.5089\\
 & \begin{tabular}{c}
SDCF 
 4L 
\end{tabular}  & 0.4014 & 0.3934 & 0.4262\\
& CNN & \textbf{0.4882} & \textbf{0.4694} & 0.5556 \\
&FCN&0.4741& 0.4666 &0.5723\\
\hline
\end{longtable}

\clearpage
\section{Financial Results for Stock Trading}
\label{appendix_b}
\noindent This section mentions the table with financial results for stock trading for all the stocks. 

\begin{longtable}[!htb]{|c|c|c|c|c|}
\caption{\textbf{Financial Results for Stock Trading }}\label{b_detailed}\\
\hline
\textbf{SYMBOL} & \textbf{Method} & 
\begin{tabular}[h]{c}
\textbf{True AR}
\end{tabular}
& 
\begin{tabular}[h]{c}
\textbf{Predicted}\\
\textbf{AR}
\end{tabular}
& 
\begin{tabular}[h]{c}
\textbf{Absolute}\\
\textbf{Difference AR}
\end{tabular}
\\
\hline
\textbf{ALKYLAMINE.BO}&\begin{tabular}{c}SDCF 1L\end{tabular}&\textbf{86.3000}&5.7500&80.5500\\
&\begin{tabular}{c}SDCF 2L\end{tabular}&&12.3000&74.0000\\
&\begin{tabular}{c}SDCF 3L\end{tabular}&&\textbf{25.9600}&\textbf{60.3400}\\
& \begin{tabular}{c}
SDCF 
 4L 
\end{tabular}  &  & 10.3700 & 75.9300\\
& CNN &  & 8.2300 & 78.0700\\
&FCN&&1.6500&84.6500\\
\hline
\end{longtable}

\clearpage
\section{Performance Analysis for SDCF vs. CNN}
\label{appendix_c}
\noindent This section mentions the table with the summary of number of stocks achieving good performance under SDCF and CNN, giving the comparative analysis of the performance between the two techniques.

\begin{ThreePartTable}
\footnotesize
\begin{longtable}{|c|c|c|c|c|c|c|c|c|c|c|c|c|c|c|}
\caption{\textbf{Comparative performance of SDCF vs. CNN in terms of \#stocks }}\label{comparison}\\
%\begin{tabular}{|c|p{6.9cm}|l|}
\hline
 \multirow{2}{*}{\textbf{Model}} & \multirow{2}{*}{\textbf{Performance}} & \multicolumn{3}{c|}{\textbf{BUY}} & \multicolumn{3}{c|}{\textbf{HOLD}} & \multicolumn{3}{c|}{\textbf{SELL}} & \multicolumn{3}{c|}{\textbf{WEIGHTED}} & \multirow{2}{*}{\textbf{AR}}\\
\cline{3-14}
& & \textbf{F1} & \textbf{P} & \textbf{R} & \textbf{F1} & \textbf{P} & \textbf{R} & \textbf{F1} & \textbf{P} & \textbf{R} & \textbf{F1} & \textbf{P} & \textbf{R} &\\
\hline
\textbf{SDCF} & \textbf{best$(>)$} & 6 & 11 & 2 &	6 &	6 &	0 &	5 & 9 &	4 &	6 &	9 &	4 & 5\\
& \textbf{equal$(=)$} & 2 & 2 & 3 & 1 & 1 & 5 & 2 & 2 & 2 & 0 & 0 &	1 &
3\\
\cline{2-15}
& \textbf{Total $>=$} & \textbf{8} & \textbf{13} & \textbf{5} & \textbf{7} & \textbf{7} & \textbf{5} & \textbf{7} & \textbf{11} & \textbf{6} & \textbf{6} & \textbf{9} & 5 & \textbf{8} \\
\cline{2-15} 
 & \textbf{next best/similar}  & 1 & 1 & 1 & 7 & 4 & 5 & 1 & 1 & 1 & 6 & 3 & 6  & 0 \\
\hline
\textbf{CNN} & \textbf{best$(>)$} & 0 & 0 & 0	& 5 & 1 & 4	& 0	& 0	& 0	& 4	& 2	& 6 & 2\\
 & \textbf{equal$(=)$} & 2 &	2 & 2 &	1 &	0 &	5 &	2 &	2 &	2 &	0 &	0 &	1 & 3
 \\
\cline{2-15} 
 & \textbf{Total $>=$} & 2 & 2 & 2 & 6 & 1 & 9 & 2 & 2 & 2 & 4 & 2 & \textbf{7} & 5 \\
\cline{2-15} 
 & \textbf{next best/similar}  & 0 & 0 & 0 & 2 & 3 & 0 & 0 & 0 & 0 & 3 & 2 & 3  & 0\\
 \hline
\end{longtable}
\begin{tablenotes}
\item \small{F1 - F1 Score, P - Precision, R - Recall}
\item \small{W - Weighted}
\item \small{AR - Annualized Returns}
\end{tablenotes}
\end{ThreePartTable}
\end{document}